\documentclass[10pt,reqno]{amsart}
\usepackage{amssymb}
\usepackage{amsbsy}
\usepackage[dvips]{color}
\usepackage{natbib}

\usepackage{amsthm}
\usepackage{amsmath}
\usepackage{amsfonts}
\usepackage{mathrsfs}
\usepackage{epsfig}
\usepackage{graphicx}

\newtheorem{thm}{Theorem}[subsection]

\newtheorem{lem}[thm]{Lemma}

\newtheorem{defn}[thm]{Definition}
\newtheorem{remark}[thm]{Remark}

\newcommand{\figdir}{figures}
\newcommand{\be}{\begin{equation}}
\newcommand{\ee}{\end{equation}} 

\begin{document}

\pagestyle{plain}

\title{Discrete Moser-Veselov Integrators for Spatial and Body Representations of Rigid Body Motions}
 
\author{Matthew Dixon\\
Department of Mathematics,\\ 
Imperial College,\\ 
London,\\
SW7 2AZ, England.\\\\
\lowercase{Matthew.dixon@imperial.ac.uk}}

\thanks{The author would like to thank his advisor, Darryl Holm, for referring him to the
work of Moser and Veselov, introducing
the discrete Clebsch approach, for useful comments on the
results and instructive advice on how to develop DMV integrators beyond the rigid body in a unified framework. The author is grateful for Jerry Marsden's
encouraging comments on a presentation of this work at the two-day conference
on Geometric Mechanics held at the University of Surrey in June, 2006.  This gratitude is extended to Colin Cotter for his assistance with the discrete-Clebsch
approach.}
\date{\today}
\maketitle
\begin{abstract}
The body and spatial representations of rigid body motion correspond, 
respectively, to the convective and spatial representations of continuum dynamics. 
With a view to developing a unified computational approach for both types of problems, 
the discrete Clebsch approach of Cotter and Holm \cite{COTTER06} for continuum mechanics 
is applied to derive (i) body and spatial representations of discrete time models of various rigid body motions
and (ii) the discrete momentum maps associated with symmetry reduction for these motions. For these problems, this paper shows that the discrete Clebsch approach yields a known class of explicit variational integrators, 
called discrete Moser-Veselov (DMV) integrators. The spatial representation
of DMV integrators are Poisson with respect to a Lie-Poisson bracket for the
semi-direct product Lie algebra. Numerical results are presented
which confirm
the conservative properties and accuracy of the numerical solutions. 
\end{abstract}

\tableofcontents

\section{Introduction}

The Hamiltonian structure of continuum mechanics in the material, inverse material, spatial and convective representations was introduced in Holm, Marsden and Ratiu \cite{HOLM86}. This work identified the body and spatial representations of rigid body motions as prototypes for the respective convective and spatial representations of continuum dynamics. Its comparison of the spatial and convective representations also put the Hamiltonian treatments of elasticity by Holm and Kupershmidt \cite{HOLM83} and by Marsden, Ratiu and Weinstein \cite{MARSDEN84} into a unified framework.
\vspace{.5cm}
\paragraph{\emph{The convective representation}} The convective and also the inverse material (augmented Eulerian) representations offer alternative descriptions of continuum models. The motivation for the convective representation of the continuum comes arose from a number of sources in the 1980s, including the study of relativistic adiabatic fluids by Holm \cite{HOLM85}, stability analysis of the coupled rigid body-beam and plate models of Krishnaprasad and Marsden \cite{KRISHNA87} and the geometrically
exact rod and plate models of Krishnaprasad, Marsden and Simo \cite{KRISHNA88}.
\vspace{.5cm}
\paragraph{\emph{Semi-direct products}} 
Holm, Marsden and Ratiu \cite{HOLM98}  derived the Euler-Poincar\'e (EP) formulation of the Eulerian fluid equations for an ideal fluid by applying symplectic reduction to Hamilton's principle for fluids.  Legendre-transforming the EP theory recovered the semidirect-product Lie-Poisson Hamiltonian theory that had been discovered and applied earlier for nonlinear stability analysis in Holm, Marsden, Ratiu and Weinstein \cite{HoMaRaWe1985}. A key step in the analysis of nonlinear stability of fluid equilibria relies on the existence of ``Casimirs'' -- quantities whose Lie-Poisson bracket vanishes with all Eulerian (spatial) fluid variables because of right-invariance of the Eulerian variables under reparameterisation of the Lagrangian labels. Because their Poisson brackets with the Hamiltonian vanish, the Casimirs are conserved quantities. We shall use this result to verify that our numerical experiments preserve the Lie-Poisson structure for the problems we consider below by explicitly showing that the values of the corresponding Casimirs are preserved.\\

\vspace{.5cm}
\paragraph{\emph{Circulation theorems}} Fluid mechanics literature widely
refers to the reparameterisation of labels as fluid parcel relabelling and attributes
the existence of the Kelvin circulation theorem for ideal flow to the application
of Noether's theorem for the particle relabelling symmetry group. Holm, Marsden and Ratiu \cite{HOLM98} showed that when advected quantities are present,
a corollary of the EP framework is a geometric form of the Kelvin circulation
theorem referred to as the Kelvin Noether theorem. In this framework,  Holm, Marsden and Ratiu \cite{HOLM86}\cite{HOLM98} further revealed the utility
of simple finite dimensional
examples, such as the heavy top, by demonstrating that they also exhibit
a Kelvin Noether theorem. This theorem together with the EP equations form an essential
ingredient in geometric models of idealised continua.

\vspace{.5cm} 
\paragraph{\emph{Variational integrators}} Geometric numerical methods seek
to transfer these powerful concepts in geometric mechanics to computational
models. In 1991, the pioneering work of Moser and Veselov \cite{MOSER91} revealed integrable classical mechanical systems which have integrable discrete time counterparts. They considered the free rigid body as one example and derived a discrete
analogue to the Euler-Arnold equations for rigid body motion in the body
description. These integrators, referred to as discrete Moser-Veselov (DMV) integrators, conserve the rigid body energy to an arbitrary order of the time step size and angular momentum to numerical round off.

Moser's and Veselov's key step was to form a discrete Hamilton's action principle
and then derive ``variational integrators'' which preserve the variational structure. Although the number of contributions to this approach is too extensive to list here, the reader may follow some important aspects of its development in Bobenko
and Suris \cite{BOBENKO99},  Marsden, Pekarsky, and Shkoller \cite{MARSDEN99}, Marsden and West \cite{MARSDEN01}, McLachlan and Scovel \cite{MCLACHLAN95},
Wendlandt and Marsden \cite{WENDLANDT97} and, more recently, Leok, Marsden and Weinstein \cite{LEOK04} who provide a differential geometric foundation for variational integrators applied to mechanical systems. The number of numerical studies supporting the theory appears less extensive, however,
largely due to the absence of a unified computational framework for deriving
practical variational integrators.

\vspace{.5cm}
\paragraph{\emph{Practical integrators}} 
DMV integrators are explicit. Cardoso and Leite \cite{CARDOSO01} 
cast the expression for the discrete angular momentum of Moser's and Veselov's
rigid body into an algebraic Ricatti equation and solved it by Schur decomposing
the Hamiltonian matrix. This gives a nearly explicit DMV algorithm, except
for the iterative
Schur decomposition. McLachlan and Zanna
\cite{MCLACHLAN05} recently demonstrated how to avoid the costly computation of the Schur form by using
an explicit spectral decomposition of the Hamiltonian instead. The Hamiltonian
can be decomposed in this way whenever its characteristic polynomial can be solved analytically.  The simple models considered here did not require this optimisation step.

\vspace{.5cm}
\paragraph{\emph{Aims}}
In this paper, we seek
to develop a unified practical computational framework for continuum dynamics
by deriving a class of explicit variational integrators for various rigid body motions in the
body and spatial representations using the same underlying approach, referred to as the discrete Clebsch approach. The continuous time Clebsch approach
provides a systematic means of deriving the EP equations from a symmetry
reduced Hamilton's action principle and the ``momentum maps'' (see \cite{MARSDEN99a}
for an explanation of this term) associated
with this symmetry reduction. For the case of rigid body motions, we aim to show that the discrete Clebsch approach provides discrete analogues, the discrete EP equations and the corresponding
momentum maps. We also aim to assess the conservative properties and accuracy
of the DMV integrators derived in this framework through
numerical experiments presented herein.

\subsection{Approach}
 We begin in section \ref{sect:discrete} by returning to arguably the most
 celebrated model taught in mechanics, the classical free rigid body in continuous time whose description, in the body frame, can be found in an
exceptionally lucid and concise form in Marsden and Ratiu's introductory
text book on geometric mechanics \cite{MARSDEN99a}. We 
slightly modify the notation for the purposes of describing the kinematics,
symmetries and equations of motion of the free rigid body in discrete time.
We also include the spatial description of the rigid body. In each case, we apply the Clebsch
approach of Cotter and Holm \cite{COTTER06}, a discrete extension of the earlier work of Cendra and Marsden \cite{CENDRA87} and Holm and Kupershmidt\cite{HOLM83}, to both derive the basic discrete EP
equations in body variables and the discrete EP equations with an advected parameter in spatial variables. 

We also derive the right and left infinitesimal equivariant momentum maps associated with
cotangent lifted left and right actions of SO(3) on the canonical and augmented
cotangent bundle respectively. We finally demonstrate the conservative properties
and accuracy of the discrete EP equations
 through numerical experiments of free rigid bodies, heavy tops and coupled rigid
bodies in section \ref{sect:num_exp}.

Appendix A provides a summary of the main features of
the spatial and body representations of each of the models that we consider, together with a comparison of their continuous time counterparts. The geometric form of
the discrete and continuous time models, derived from a Clebsch approach,
are remarkably similar. For completeness, appendix B provides the
spatial version of McLachlan's and Zanna's (unoptimised) DMV algorithm \cite{MCLACHLAN05}. We now state the main results of this paper.

\newpage
\subsection{Summary of main results}

\paragraph{\emph{Discrete Clebsch approach}} We show how
this discrete Clebsch approach not only recovers the body representation of the basic discrete EP equations of \cite{MOSER91} but also yields a spatial representation of the EP equations
with an advected parameter which correspond to the Lie-Poisson equations on the dual space of a semidirect product lie algebra, discovered
by Bobenko and Suris \cite{BOBENKO99} for a more general class of problems when the phase
space is $G\times G$. Bobenko and Suris also showed that the Lax representation
in the spatial frame is gauge invariant
to its representation in the body frame. Bobenko and Suris's discovery of the semi-direct product
in discrete time enables the transfer of the significant work
of Holm, Marsden and Ratiu on semi-direct products in continua to the discrete time case. Prior to their paper, the relation between EP equations with an
advected parameter and Lie-Poisson Hamiltonian systems required invertibility
of the partial Legendre transform. This work overcame this restriction by
developing the EP theory entirely within the Lagrangian framework.

\vspace{.5cm}
\paragraph{\emph{Momentum maps}} We derive right
and left momentum
maps for the respective left and right reductions of the Lagrangian by SO(3).
The right
momentum map is associated with the cotangent lifted left action of SO(3) on the approximated cotangent bundle
and its image is the spatial angular momentum, which is an element of $\mathfrak{so}(3)^*$.
Left invariance of the action principle implies conservation of spatial angular momentum $m$. The
image of the left momentum map is the body angular momentum $M$ which is only conserved when the
action principle is right invariant. This is the limiting case when one or more of the moments of inertia are equal
and is not considered in this paper.
\vspace{.5cm}
\paragraph{\emph{Conserved quantities}}  
We prove that both representations of the DMV integrator conserve the spatial angular
momentum and demonstrate through numerical experiment that the DMV algorithms conserve spatial
angular momentum to numerical round off. We also demonstrate the correct
computation of the Casimirs $||M||^2_2$ and $\{||I||_2$,~$det(I)\}$ of the Lie-Poisson bracket for the respective
body and spatial representations and
that the numerical solution conserves energy to an order of the time step. 
\vspace{.5cm}
\paragraph{\emph{The heavy top}} We apply the same approach to the classical
model of the heavy top. For a detailed treatment of the motion of the heavy top we refer the reader
to the work of Lewis, Simo and Marsden \cite{LEWIS92}. They consider three
types of heavy tops, the asymmetric, the tilted and sleeping Lagrangian tops and extend
the classical studies of the heavy top by using the reduced energy momentum
to derive relative equilibria.  

A result which seeks
numerical validation is the observation that stable branches of steadily precessing Lagrangian tops bifurcate from the branch of sleeping Lagrange lops throughout the range of angular
velocities for which the sleeping top is stable. We do not address the extension of these results
to the discrete Lagrangian framework but instead present some geometric properties in the discrete framework (in both representations) together with numerical results
of the motion of the heavy top in the body representation only. Appendix
A provides a description
of the geometric properties of the body and spatial representations of the
heavy top motion in discrete and continuous time.

\vspace{.5cm}
\paragraph{\emph{The coupled rigid body}}  We
apply the discrete Clebsch approach to the classical coupled rigid body.
We begin by presenting a slightly revised description (in the frame of body 1) of the $\mathbb{R}^3$
reduced coupled rigid body
which is clearly described in the thesis of Patrick \cite{PATRICK89}
and the concise paper by Grossman, Krishnaprasad and Marsden \cite{GROSSMAN88}.
The last table of appendix A provides a spatial description of the continuous time problem together with a corresponding description in discrete time.
We also hope that this description of the discrete coupled rigid body will
be extended to form a discrete analogue of the engineering related work on
coupled rigid bodies and geometrically exact rods by Simo, Posbergh and Marsden \cite{SIMO90} and the study of the dynamics and stability of the coupled rigid
bodies by Sreenath, Krishnaprasad and Marsden
\cite{SREENATH88}. We provide some results from numerical
experiments of the coupled rigid body as viewed in the frame of body 1.

\subsection{Important related works}
Similar approaches derive discrete equations of rigid body motion for optimal control problems.
Bloch, Crouch, Marsden and Ratiu \cite{BLOCH02}, for example, derive the symmetrised rigid
body equations by introducing optimality constraints in the action principle. We distinguish our approach from theirs in two ways. Firstly,
although they consider the rigid body motion as an optimal control problem
with an associated constrained action principle, they do not identify the
constraints as Clebsch variables and derive the momentum maps. Secondly,
they present left and right trivialisations of $T^*SO(3)$ where
as we present body and spatial representations of a left SO(3) action invariant Lagrangian only. The authors make this point when distinguishing
their approach from that of Holm and Kupershmidt \cite{HOLM83}. We use the expression for the (left) momentum map to prove that the flow on the cotangent bundle preserves spatial angular momentum and derive
the equations of motion. 

The subsequent presentation of
the discrete equations of motion and momentum maps for the heavy top appears unique in so far as the only citation found on geometric integrators
for the heavy top, by Celledoni and S\"afstr\"om \cite{CELLEDONI05}, does not consider DMV integrators nor the spatial representation. Work on practical
aspects of geometric integrators for the discrete coupled
rigid body problem is not cited in the literature although Patrick \cite{PATRICK89} references
several of his own Maple scripts for computing various properties of the coupled rigid
body.

\section{The Free Rigid Body} \label{sect:discrete}

In this section, we slightly modify the description of the free
rigid body given in chapter 15 of \cite{MARSDEN99a} as a discrete time problem
in both the body and the spatial representations.
Consider a free rigid body as a solid body occupying a \emph{reference configuration} $\mathcal{B}\in \mathbb{R}^3$
which is free to move in the container $C=\mathbb{R}^3$ by rotations about
a fixed point. Material points $\ell\in\mathcal{B}$ are position
vectors whose components, relative to a fixed orthonormal basis $(\mathbf{E}_1,\mathbf{E}_2,\mathbf{E}_3)$ in $\mathcal{B}$, are the \emph{material coordinates}. A \emph{configuration} $\mathfrak{C}$ of $\mathcal{B}$ is a continuous, invertible and orientation preserving map $\phi:\mathcal{B}\rightarrow C$ from material points to \emph{spatial}
points in the container. The spatial points are position vectors whose components,
relative to $(\mathbf{e}_1,\mathbf{e}_2,\mathbf{e}_3)$, the right-handed orthonormal basis of $C,$ are
\emph{spatial coordinates}. 

Discrete time dependent families of configurations of $\mathcal{B}$ are referred to
as a \emph{discrete time motion} of $\mathcal{B}$ and are written in terms of the \emph{forward map} $\mathbf{x}=\psi(\mathbf{\ell},t_k)=\psi_{t_k}(\mathbf{\ell}),~~k\in \mathbb{Z}^+$, which in addition to having the properties
of the configuration also satisfies $\psi(\mathbf{\ell},t_0)=\psi_{t_0}(\mathbf{\ell})=\mathbf{\ell}$.

 This
last property together with rigidity of the body and continuity of the motion implies that the configuration of $\mathcal{B}$ may be identified
with $SO(3)$ and the forward map is written
as 
\be
\mathbf{x}=\psi_{k}(\ell)=\Lambda_k\mathbf{\ell},~~\Lambda_k\in SO(3),
\ee
where, for notational convenience, $\psi_k:=\psi_{t_k}$ and $\Lambda_k:=\Lambda(t_k)$,
represents the attitude of the body.
Put simply, the forward map is the position
of a label $\mathbf{\ell}$ in the container. The value of the label is 
its position in the container at time $t_0$. 

The \emph{body} coordinates of a material position vector are its components
relative to a \emph{time-dependent} basis $(\mathbf{\xi}_1,\mathbf{\xi}_2,\mathbf{\xi}_3)(t_k)$
which is defined by $\mathbf{\xi}_i(t_k)=\Lambda_{k} E_i,~~i:=1\rightarrow 3$ and
hence is attached to the rigid body that rotates about the origin of C. 

\subsection{Discrete velocities}

The image of the \emph{backward or inverse map} $\mathbf{\ell}=\psi^{-1}_k(\mathbf{x})$ is the label
whose position at time $t_0=0$ is $\mathbf{x}$. With the configuration identified
as $SO(3)$, this is given by

\be
\mathbf{\ell}=\psi^{-1}_k(\mathbf{x})=\Lambda_k^{-1}\mathbf{x}.
\ee

It follows that label positions at consecutive times $t_{k-1}$ and $t_{k}$ are related by the composite of forward and backward maps which are group
products 
\be
\psi_{k}(\mathbf{\ell})=\psi_k\circ\psi^{-1}_{k-1}\circ\psi_{k-1}(\mathbf{\ell})=\Lambda_k \Lambda_{k-1}^{-1}\Lambda_{k-1}(\mathbf{\ell}).
\ee

The product of a matrix and an inverse matrix at consecutive times is referred to as a \emph{discrete velocity}. Just as the (continuous time) spatial and body velocities are right and left invariant respectively, so
too are their discrete counterparts.

The spatial discrete angular velocity $\omega_{k+1}$ and body discrete angular
velocity $\Omega_{k+1}$ respectively satisfy the reconstruction formulae
\be
\begin{split}
\Lambda_{k+1}&=\omega_{k+1}\Lambda_{k},\\
\Lambda_{k+1}&=\Lambda_{k}\Omega_{k+1}.\\
\end{split}
\ee

\noindent The two discrete velocities at time $t_{k+1}$ are related by 

\be
\Omega_{k+1}=\Lambda_{k}^T w_{k+1} \Lambda_{k}=Ad_{\Lambda_{k}^T}w_{k+1}.
\ee
\begin{remark}
The reconstruction formulae conserve labels. This statement is confirmed
by considering the spatial reconstruction formula

\be
\begin{split}
X_{k+1}&=\omega_{k+1} X_k\\
\Lambda_{k+1}^T X_{k+1}&=\Lambda_k^T X_k\\
\ell_{k+1}&=\ell_k.\\
\end{split}
\ee
\end{remark}

With the spatial and body discrete velocities defined, we now consider the
geometric mechanics of rigid body motion in discrete time.

\section{Discrete Constrained Variational Principle}\label{sect:dvp}

Moser and Veselov \cite{MOSER91} consider a Lagrangian $L:G\times G \rightarrow \mathbb{R}$ which is a smooth map defined as

\begin{equation}
\label{eqn:disc_lag_material}
L(\Lambda_k,\Lambda_{k+1})=Tr(\Lambda_kI_0\Lambda_{k+1}^T)
-Tr\left(\Theta_{k+1}(\Lambda_{k+1}\Lambda_{k+1}^T-I_d)\right).
\end{equation}

The constant inertia matrix $I_0\in V^*=S_2(\mathbb{R}^3)$ is a matrix
of the positive symmetric covariant two-tensors on $\mathbb{R}^3$. We identify $V$
with $V^*$ using the metric $Tr(A^TB)$ so that $V=S^2(\mathbb{R}^3)$
is the $3\times3$ matrix of symmetric covariant two-tensors on $\mathbb{R}^3$ dual to $S_2(\mathbb{R}^3)$. The pairings in the above Lagrangian are therefore between $V$ and $V^*$. 

We denote the projection of elements $g\in G$ onto $S^2(\mathbb{R}^3)$ as $sym(g)$, where $sym()$ is a projection operator defined as $sym(g):=\frac{1}{2}(g+g^T)$.
The symmetric Lagrange matrix multiplier $\Theta_{k+1}\in
V^*$ enforces orthogonality of $\Lambda_k$\footnote{This constraint is required to derive the basic discrete EP equations on $SO(3)$ and not on $S^2(\mathbb{R}^3)$.}.

The problem is to find the sequence $\{\Lambda_k,\Lambda_{k+1}\}$ which satisfies the
discrete stationary action principle \footnote{Holm, Marsden and Ratiu \cite{HOLM98}
point out that this is not strictly a variational principle (because the variations
of the dynamical variables are constrained) but a Lagrange D'Alembert principle.}
\be
0=\delta S_{d}=\sum_k L(\Lambda_k,\Lambda_{k+1}).
\ee
A necessary condition for extremising this functional is that $\{\Lambda_k,\Lambda_{k+1}\}$ satisfy
the discrete Euler-Lagrange equations. We shall now derive the solution in the spatial and body representations.

\section{Symmetry Reduction}\label{sect:sym_red}

\paragraph{\emph{Body representation}} The discrete differential geometric formulation of the DMV system
is given in \cite{LEOK04}. The principal G-Bundle $(G\times G,G,\pi)$ together
with the
natural projection $\pi:G\times G \rightarrow G\times G/G$ furnish the description
of discrete symmetry reduction to the body representation.

\begin{figure}[!ht]
\begin{center}
\includegraphics[angle=0.0,width=0.6\textwidth]{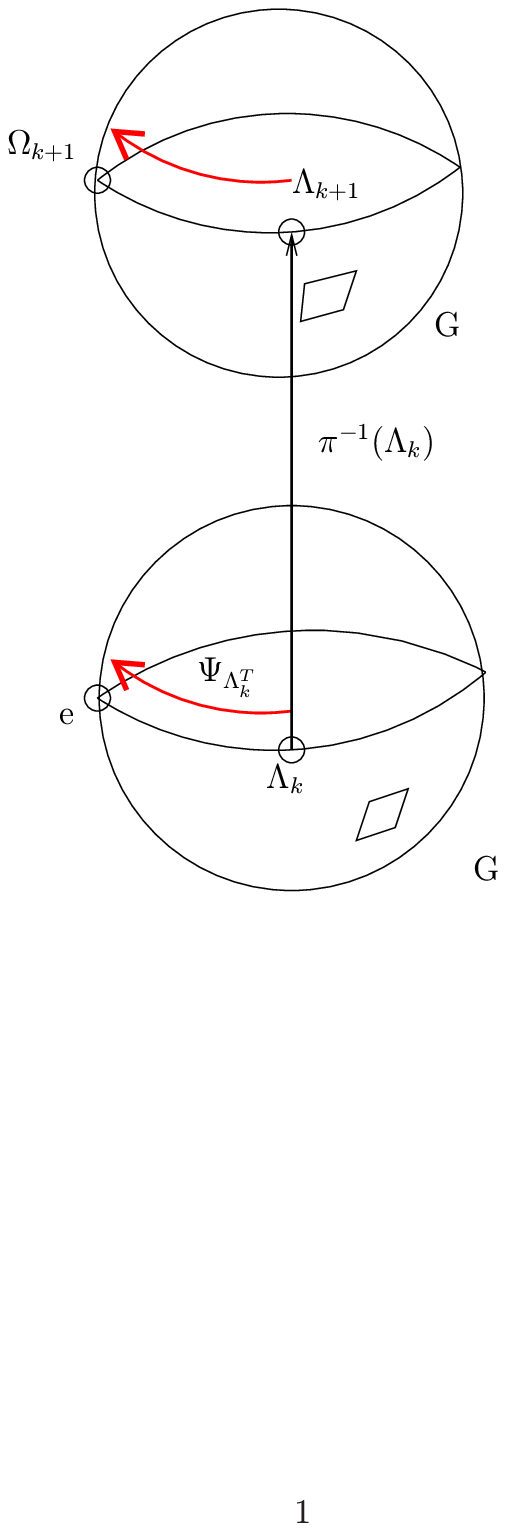}\\
\end{center}
\caption{\scriptsize{The principal G-Bundle upon which the discrete Lagrangian is defined.
The two curved arrows represent the diagonal action of $\Lambda^T_k$ on $(\Lambda_k,\Lambda_{k+1})$.
The vertical arrow represents the fiber over $\Lambda_k$.}}
\end{figure}

\begin{defn}
The (left) \textbf{diagonal action} of $G$ on $G\times G$ is defined as
$\Psi: G\times(G\times G) \rightarrow G\times G ~~|~~\Psi(f,(g,h))=
f\cdot(g,h)=(fg,fh)$.
\end{defn}

The discrete Lagrangian in equation \ref{eqn:disc_lag_material} is invariant under
the (left) action of $\Psi$. As depicted by the two curved arrows in figure 1,
one may reduce the Lagrangian on the G-bundle by this action to obtain the
reduced Lagrangian $L:G\rightarrow\mathbb{R}$ given in body variables by

\be
L(\Omega_{k+1})=Tr(sym(\Omega_{k+1})I_0)
-Tr\left(\Theta_{k+1}(\Omega_{k+1}\Omega_{k+1}^T-I_d)\right).
\ee

\vspace{0.5cm}
\paragraph{\emph{Spatial representation}} To reduce to spatial variables, we must firstly modify the definition of the Lagrangian for reasons
which are give in section 6. We consider the Lagrangian defined on the \emph{augmented} space,
\be 
L:SO(3)\times SO(3)\times V^* \rightarrow \mathbb{R},
\ee
\noindent given by
\be \label{eqn:lgr_augm}
L(\Lambda_k,\Lambda_{k+1},I_0)=Tr(\Lambda_kI_0\Lambda_{k+1}^T)
-Tr\left(\Theta_{k+1}(\Lambda_{k+1}\Lambda_{k+1}^T-I_d)\right).
\ee

We must define how SO(3) acts on $V^*$. Let $\Phi^*(g)\cdot I,~I\in V^*,~g\in G= SO(3)$ be the left group anti-representation on $V^*$. This action defines an orbit of the inertia matrix in
$V^*$, which is a dynamical variable in the spatial frame. 
 
The left action $\Phi^*(g)$ on $V^*$ is given by
\be
\Phi^*:G\times V^* \rightarrow V^* ~~~|~~~\Phi^*(g)\cdot I=gIg^{-1}.
\ee
The corresponding
right action is given by
\be
\Phi^*:V^*\times G \rightarrow V^* ~~~|~~~I\cdot\Phi^*(g)=g^{-1}Ig,
\ee
which is equal to the left anti-representation $\Phi^*(g^{-1})\cdot I$.

\begin{defn}
The (right) \textbf{augmented diagonal action} of $G$ on $G\times G\times
V^*$ is defined as
$\Psi': (G\times G\times V^*)\times G \rightarrow G\times G\times V^* ~~|~~\Psi'((g,h,a),f)=
(g,h,a)\cdot f=(gf,hf,a\cdot\Phi^*(f))$.
\end{defn}

This discrete Lagrangian, which is now a function of $I_0$, is invariant under the (right) action $\Psi'$ of $g\in SO(3)$

\be
L(\Lambda_kg,\Lambda_{k+1}g,g^TI_0g)=L(\Lambda_k,\Lambda_{k+1},I_0).
\ee

Reduction of the Lagrangian by the augmented diagonal action gives the reduced Lagrangian $L:G\times V^* \rightarrow \mathbb{R}$ given in spatial variables as

\be
L(\omega_{k+1},I_k)=Tr(sym(\omega_{k+1})I_k) -Tr\left(\Theta_{k+1}(\omega_{k+1}\omega_{k+1}^T-I_d)\right).
\ee

\section{Clebsch Potentials and Momentum Maps}\label{sect:clebsch}

The symmetry reductions have an associated momentum map which may be derived
by adopting the discrete Clebsch approach of Cotter and Holm \cite{COTTER06}.
An associated or augmented Lagrangian is defined by adding a Clebsch potential to enforce
the reconstruction formula. We will consider the approach in body and spatial
representations.

\subsection{The body representation}
In body
variables the augmented Lagrangian is given by
\be
L':=Tr\left(I_0sym(\Omega_{k+1})\right) -\frac{Tr}{2}\left(P_{k+1}^T(\Lambda_{k+1}-\Lambda_k\Omega_{k+1})\right)-Tr\left(\Theta_{k+1}(\Omega_{k+1}\Omega_{k+1}^T- I_d)\right).
\ee

The stationary constrained action
principle gives Clebsch relations which include the discrete Euler-Lagrange equation

\be
\nabla_{\Lambda_{k}} L'(\Lambda_{k-1},\Lambda_{k}) + \nabla_{\Lambda_{k}}L'(\Lambda_{k},\Lambda_{k+1})=0.\\
\ee
Evaluation of
the derivatives of $L'$ and subsequent rearrangement gives an evolution
equation for the Lagrange multiplier $P_k$

\be
P_{k+1}=\omega_{k+1}P_k.
\ee
This equation together with the discrete evolution equation for $\Lambda_{k}$
describe the discrete flow on the cotangent bundle and are referred to as the discrete \textit{symmetrised equations} and are analogous to the continuous
symmetrised equations.

The Clebsch relation paired with the variation $\delta\Omega_{k+1}$
in the discrete action principle gives the expression
\be
J^R_{k+1}=\text{skew}(\Lambda_k^T P_k),
\ee
which satisfies the definition of the momentum map for cotangent lifted left
actions of SO(3)

\be
\begin{split}
\langle J^R_{k+1},\zeta\rangle&=\langle P_k, \mathfrak{L}_{\zeta}\Lambda_k\rangle\\
&=\langle P_{k}\diamond\Lambda_{k}, \zeta \rangle,
\end{split}
\ee
where the bilinear operation
$\diamond:V\times V^*\rightarrow \mathfrak{g}^*$ is defined in \cite{HOLM98}
and $\langle A,B\rangle=-\frac{1}{2}Tr(A^TB)$ for any matrices $A,B\in V$.

It is well known that momentum maps for cotangent lifted actions are equivariant and consequently Poisson. The image of this map is the body angular momentum $M_{k+1}:=I_0\Omega_{k+1}^T-\Omega_{k+1}I_0$.

Substitution of the symmetrised equations for the discrete flow on the cotangent bundle

\be
\begin{split}
P_{k+1}&=P_k\Omega_{k+1},\\
\Lambda_{k+1}&=\Lambda_k\Omega_{k+1},
\end{split}
\ee

\noindent into the right momentum map gives the discrete basic EP equation
defined on $SO(3)$ for rigid body motion in the body frame

\be
M_{k+1}=Ad^*_{\Omega_{k}^T}M_k.
\ee

\subsection{The spatial representation}

We now derive the equations
of motion in the spatial representation by applying the discrete Clebsch approach. In spatial variables, the Clebsch modified discrete Lagrangian is given by

\be
\begin{split}
L'&:=Tr\left(I_k sym(\omega_{k+1})\right) -\frac{Tr}{2}(P_{k+1}^T(\Lambda_{k+1}-\omega_{k+1}\Lambda_k)\\& - \frac{Tr}{2}(J_{k+1}(I_{k+1}
-\omega_{k+1}I_k\omega_{k+1}^T)) -Tr \left(\Theta_{k+1}(\omega_{k+1}\omega^T_{k+1}
- I_d)\right).
\end{split}
\ee

The stationary constrained action
principle gives Clebsch relations which are the discrete Euler-Lagrange equations

\be
\begin{split}
\nabla_{\Lambda_{k}}L'(\Lambda_{k-1},\Lambda_{k}) + \nabla_{\Lambda_{k}}L'(\Lambda_{k},\Lambda_{k+1})&=0,\\
\nabla_{I_{k}}L'(I_{k-1},I_{k}) + \nabla_{I_{k}}L'(I_{k},I_{k+1})&=0.\\
\end{split}
\ee

Evaluation of
the derivatives of $L'$ and subsequent rearrangement gives, respectively,

\be
\begin{split}
P_{k+1}&=\omega_{k+1}P_k,\\
J_{k+1}&=\omega_{k+1}\underbrace{(-sym(\omega_k) + J_{k})}_{G_k}\omega_{k+1}^T.
\end{split}
\ee

Additionally, the Clebsch relation paired with $\delta \omega_{k+1}$ gives
\be
I_{k+1}\omega_{k+1}+ A_k P_{k+1}^T\omega_{k+1} +I_k \omega_{k+1}^T J_{k+1}\omega_{k+1}=\Theta_{k+1}. \ee

Using the symmetry property of $\Theta_k$, the equations $P_k^T=P_{k+1}^T\omega_{k+1}$ and
$G_k=\omega_{k+1}^T J_{k+1}\omega_{k+1}$ gives an expression for $J^L_{k+1}$
\be
J^L_{k+1}=\text{skew}(P_k A_k^T) 
+[G_k,I_k],\\
\ee

\noindent which satisfies the definition of the momentum map for cotangent lifted left
actions of SO(3)
\be
\begin{split}
\langle J^L_{k+1},\zeta\rangle&=\langle P_k, \mathfrak{L}_{\zeta}\Lambda_k\rangle
+ \langle G_k, \mathfrak{L}_{\zeta}I_k\rangle\\
&=\langle P_{k}\diamond\Lambda_{k} + G_k\diamond I_k, \zeta \rangle.
\end{split}
\ee

The image of $J_L^{k+1}$ is the spatial angular momentum
$m_{k+1}:=I_{k+1}\omega_{k+1} -\omega_{k+1}^T I_{k+1}$. The spatial representation of the discrete EP equations with
an advected parameter are

\begin{eqnarray} \label{eqn:dep_spatial}
m_{k+1}&=&Ad^*_{\omega_k^T}m_k - \nabla_{I_k}L\diamond I_k,
\\
I_{k+1}&=&\omega_{k+1}I_{k}\omega_{k+1}^T.
\end{eqnarray} 

\begin{lem}
The spatial angular momentum is a conserved quantity.
\end{lem}

\begin{proof}
Substituting the symmetrised equations for $P_k$ and $\Lambda_k$ into the expression for the left momentum map gives
\be
\begin{split}
m_{k+1}&=\omega_k (P_{k-1} A_{k-1}^T -A_{k-1} P_{k-1}^T) \omega_k^T -[I_k,G_k]\\
&=\omega_k m_k\omega_k^T + \omega_k[I_{k-1},G_{k-1}]\omega_k^T -[I_k,G_k]\\
\end{split}
\ee

Using $G_k=\omega_{k}G_{k-1}\omega_{k}^T -sym(\omega_k)$
and $I_k=\omega_{k}I_{k-1}\omega_{k}^T$ gives

\be
\begin{split}
m_{k+1}&=\omega_k m_k\omega_k^T +[I_k,\omega_k+\omega_k^T]\\
&=\omega_k m_k\omega_k^T + I_k \omega_k -\omega_k^T I_k - \omega_k I_k +
I_k\omega_k^T\\
&=\omega_k m_k\omega_k^T + m_k - \omega_k(I_k\omega_k - \omega_k^T I_k)\omega_k^T\\
&=m_k.
\end{split}
\ee
\end{proof}

So for conservation of spatial angular momentum, the coAdjoint orbits of the action of SO(3) on $\mathfrak{so}(3)^*$ must take the form
\be
Ad^*_{\omega_k^T} m_0=m_0 +\nabla_{I_{k}} L \diamond I_k.
\ee

\section{Poisson Brackets and Semidirect Products}

Bobenko and Suris \cite{BOBENKO99} show that the right reduced discrete EP equations are Lie-Poisson w.r.t. a semidirect product lie algebra. We show how the
spatial representation of the discrete EP equations for the
rigid body, given above, are related to Bobenko's and Suris's result, which
is generalised to a class of systems in which the heavy
top is one example. We will firstly present the Lie-Poisson bracket for the
rotating rigid body in the spatial frame by excluding the centre of mass
vector $\chi$ from the study of the
heavy top in the spatial frame performed by Holm, Marsden and Ratiu \cite{HOLM86}.

Unless otherwise stated, we omit the time index subscripts for ease of notation.
We point out a few minor differences in our notation with Bobenko's and Suris's.
The first point is that Bobenko and Suris consider a dynamical
variable $p\in V$, in the linear space in which the group is represented, whereas we consider a dynamical variable
$I\in V^*$, in the dual of this linear space. 
Our notation is consistent
with the work by Holm, Marsden and Ratiu on semidirect products in the EP
description of the continuum \cite{HOLM98}.

We extend the definitions given in section 4. The corresponding left anti-representation of the lie algebra on $V^*$ is
\be
\Phi^*:\mathfrak{g}\times V^*\rightarrow V^* ~~~|~~~\Phi^*(\zeta)\cdot I=[\zeta,I],~\zeta
\in \mathfrak{g}=\mathfrak{so}(3),
\ee
with the right anti-representation following from the definition of the right
group action on $V^*$
\be
\Phi^*: V^*\times\mathfrak{g}\rightarrow V^* ~~~|~~~I\cdot\Phi^*(\zeta) I=[I,\zeta,],~\zeta
\in \mathfrak{g},
\ee

\noindent where it follows that 
\be \label{eqn:interchange}
\Phi^*(\zeta)\cdot I=-\Phi^*(\zeta^T)\cdot I=-I\cdot\Phi^*(\zeta).
\ee

Unlike the material representation of the general class of Lagrangians on $G\times G$ that Bobenko and Suris consider, for any group $G$, the material representation of the Lagrangian for the rigid body on $SO(3)\times
SO(3)$ is \emph{not} invariant under the diagonal right action of $G=SO(3)$.
The only exception to this occurs when one or more of the principal moments of inertia
are equal to each other for which the isotropy group $G_{I_0}$ for $I_0$  is the subgroup
$SO(2,1)$ or $SO(3)$ respectively. We do not consider this special case and instead consider the Lagrangian defined on the augmented space, given in equation \ref{eqn:lgr_augm}, which is invariant under the augmented diagonal right action
defined in section 4.

The discrete EP equation with an advected parameter $I_k$ given in equation
\ref{eqn:dep_spatial} is defined on $G\times V^*$. Under the assumption
of invertibility of the partial "Legendre transformation"
\be
(\omega_{k+1},I_{k})\in G\times V^*\rightarrow (m_{k+1},I_{k+1})\in\mathfrak{g}^*\times V^*,
\ee
given by Bobenko and Suris (equation 4.19, \cite{BOBENKO99}), equation
\ref{eqn:dep_spatial} defines the smooth map
\be \label{eqn:map}
(m_k,I_k)\mapsto (m_{k+1},I_{k+1})~ \in \mathfrak{g}^*\times V^*.\\
\ee 

This map is Poisson  w.r.t. to the $(\pm)$ Lie-Poisson brackets on the dual of the semi-direct
product algebra $[\mathfrak{s}^*=\mathfrak{g}^*\circledS V^*]_{\pm}$ which have the form
\be \label{eqn:LPB}
\begin{split}
\{F_1,F_2\}_{\pm}(m,I)&=-\frac{1}{2}Tr\{\pm m^T[\nabla_m F_1,\nabla_m F_2]\\ 
&\mp I\left(\nabla_I F_2\cdot\Phi(\nabla_m F_1)
-\nabla_I F_1\cdot\Phi(\nabla_m F_2)\right)\},
\end{split}
\ee

\noindent where $\Phi$ is the right representation of $\mathfrak{g}$ in $V$ which is related
to the anti-representation by the expression given in (equation 3.16, \cite{BOBENKO99}). These are the $(\pm)$ Lie-Poisson brackets on $\mathfrak{s}^*$ for the right representation $\Phi$ of $\mathfrak{g}$ on $V$ given by Holm, Marsden
and Ratiu (section 5.5, \cite{HOLM86}) for the spatial representation of the heavy top (in the absence
of a gravitational potential). The same authors also
give these brackets for the more general case when $G$ is any group which acts on $V^*$ from the right in (equation 2.14, \cite{HOLM98}).

The (-) Lie-Poisson bracket in equation \ref{eqn:LPB} is
the same bracket defined by Bobenko and Suris in (equation 4.20, \cite{BOBENKO99})
and so we conjecture that the proof (page 12, \cite{BOBENKO99}) of the Poisson property of the
map given in equation \ref{eqn:map} applies here. Note that they state the (-) Lie-Poisson bracket of the
semi-direct product lie algebra
corresponding to minus the left anti-representation
of $G$ on $V^*$, whereas we state the form of the $(\pm)$ brackets corresponding to
the right representation of $SO(3)$ on $V$. The relationship between the representations given in equation \ref{eqn:interchange} permits the simple interchange between the bracket corresponding
to the (-) left and right representations. The right representation is a
prototype for idealised fluids.

We close this section by commenting on the Casimirs of this Lie-Poisson bracket
which are less well-known than for the standard rigid body bracket. We firstly express the Lie-Poisson bracket in a more concise form
\be
\{F_1,F_2\}_{\pm}(m,I)=\pm\langle m,[\nabla_m F_1,\nabla_m F_2]\rangle \pm \langle I \diamond \nabla_I F_2, \nabla_m F_1\rangle
\mp\langle I\diamond \nabla_I F_1, \nabla_m F_2\rangle.
\ee

Holm, Marsden and Ratiu \cite{HOLM86} show that the Casimir functions on the Poisson manifold $[\mathfrak{s}^*]_{\pm}$
are the set of functions invariant under the coAdjoint action of the Lie group $S=SO(3)\circledS
V^*$. For the right representation of $\Lambda$ on $V^*$, the definition of this action, given in a general form in (equation 2.09, \cite{HOLM98}),
is

\be
(\Lambda,J)(\zeta,I)=\langle Ad^*_{\Lambda^T}\zeta +  J\cdot\Phi(\Lambda^T)\diamond I\cdot\Phi(\Lambda^T), I\cdot\Phi(\Lambda^T)\rangle.
\ee

It follows that invariants of $I$ under
conjugation by $\Lambda$, such as $det(I)$ and $||I||_2$ are invariant under this coAdjoint
action and are subsequently Casimirs of the Lie-Poisson bracket.

\section{The Heavy Top}\label{sect:top}

In this section, we extend the main results of the spatial and body representations
of the rigid body, presented thus far, to the heavy top.  We consider the kinematics and symmetries of the tilted Lagrange top, described in the body frame by
the orientation $\mathbf{\Gamma}_k$
of the vertical axis $\hat{z}$ and the body angular velocity (See figure (2))
and in the spatial frame by the position of center of mass $\chi_k$ relative to the support, the inertia tensor and the spatial angular velocity (see
figure (3)).

The Lagrangian top is a
special case of the heavy top which has a symmetric inertia matrix and its
centre of mass lies on its axis of symmetry. The motion of the tilted Lagrangian
top is distinct
from the sleeping top in that it both spins about its axis of symmetry and
precesses about each of the spatial axes, maintaining a positive vertical
coordinate. The axis of symmetry of the sleeping top remains parallel
to the gravitational vector $-g\hat{z}$, however.

\subsection{The body representation}
$\quad$\\
\begin{figure}[!ht]
\begin{center}
\includegraphics[angle=0.0,width=0.8\textwidth]{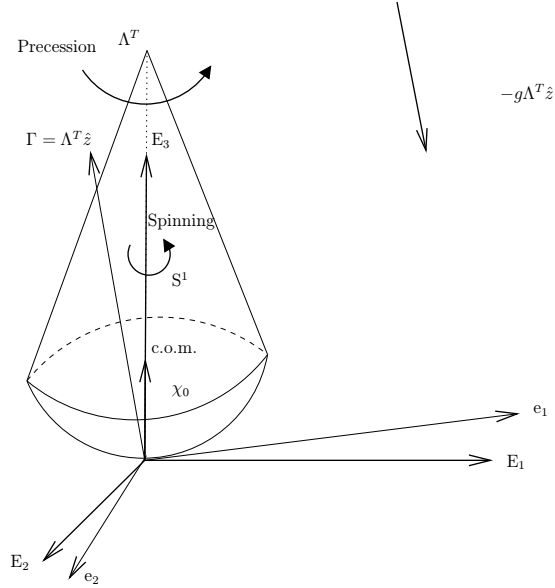}\\
\end{center}
\caption{\scriptsize{The heavy top as viewed in the body frame. The heavy body is attached to the spatial frame at an arbitrary point (in this diagram this point is the origin). The motion is composed of two components, precession
and spinning. The unit vector $\hat{z}$ , representing the direction of gravity,
rotates about each axis of the heavy top with body angular velocity $\Omega$.
Spatial angular momentum is only conserved for motions purely about $E_3$,
however. The body frame also spins
about its centreline axis but this is only observable in the spatial frame. The vector $\chi_0$ from the point of support to the centre of
mass of the top (c.o.m.), which lies on the centreline of top,  remains fixed in the body frame.}}
\end{figure}

It is well known that the motion of the heavy top breaks the symmetry of
the Lagrangian when
precession modifies the gravitational potential of the top or its
spin is not purely about its centreline.  The symmetry group
for the heavy (Lagrangian) top is thus $S^1\times S^1$. From Noether's theorem
we may deduce that spatial angular momentum will only be conserved if the
group action for the motion is that of this symmetry group.

The Clebsch variable constrained Lagrangian for the discrete time heavy top motion in the
body representation is

\be
\begin{split}
L&=-\frac{Tr}{h^2}\left(I_0sym(\Omega_{k+1})\right) - mg \langle\mathbf{\Gamma}_k,\mathbf{\chi}_0\rangle   -\frac{Tr}{2}\left(P_{k+1}^T(\Lambda_{k+1}-\Lambda_k\Omega_{k+1})\right)\\
  &-\langle \mathbf{J}_{k+1},\mathbf{\Gamma}_{k+1}- \Omega_{k+1}\mathbf{\Gamma}_k\rangle
  +Tr\left(\Theta_{k+1}(\Omega_{k+1} \Omega_{k+1}^T -I_d)\right),
\end{split}
\ee
where $h$ is the fixed time interval. The (right) momentum map corresponding to the left augmented diagonal action of $SO(3)$
on $SO(3)\times SO(3)\times \mathbb{R}^3$ is

\be
\mathbf{J}^R_{k+1}=P_k\diamond\Lambda_k + \mathbf{\Gamma}_{k}\diamond\tilde{\mathbf{J}}_{k},
\ee
with 
$\tilde{\mathbf{J}}_{k}=\mathbf{J}_{k}
+mg\mathbf{\chi}_0$.

The Clebsch relations include the discrete Euler-Lagrange equations

\be
\begin{split}
\nabla_{\Lambda_{k}}L'(\Lambda_{k-1},\Lambda_{k}) + \nabla_{\Lambda_{k}}L'(\Lambda_{k},\Lambda_{k+1})&=0,\\
\nabla_{\Gamma_{k}}L'(\mathbf{\Gamma}_{k-1},\mathbf{\Gamma}_{k}) + \nabla_{\Gamma_{k}}L'(\mathbf{\Gamma}_{k},\mathbf{\Gamma}_{k+1})&=0.\\
\end{split}
\ee

 Evaluation of
the derivatives of $L'$ and subsequent rearrangement gives, respectively,

\be
\begin{split}
P_{k+1}&=P_k\Omega_{k+1},\\
\mathbf{J}_{k+1}&=\Omega_{k+1}\tilde{\mathbf{J}}_k.\\
\end{split}
\ee

Substitution of these two equations into the equation for the (right) momentum
map gives the equations of motion in the body variables

\be
\begin{split}
M_{k+1}&=Ad^*_{\Omega_k^T}M_k +mg\mathbf{\Gamma}_k\diamond\mathbf{\chi}_0,\\
\mathbf{\Gamma}_{k+1}&=\Omega_{k+1}\mathbf{\Gamma}_k,\\
\end{split}
\ee
with $M_k:=\frac{2}{h^2}skew\left((\nabla_{\Omega_{k+1}}L)^T\Omega_{k+1}\right)$.

\subsection{The spatial representation}
$\quad$\\
\begin{figure}[!ht]
\begin{center}
\includegraphics[angle=0.0,width=0.7\textwidth]{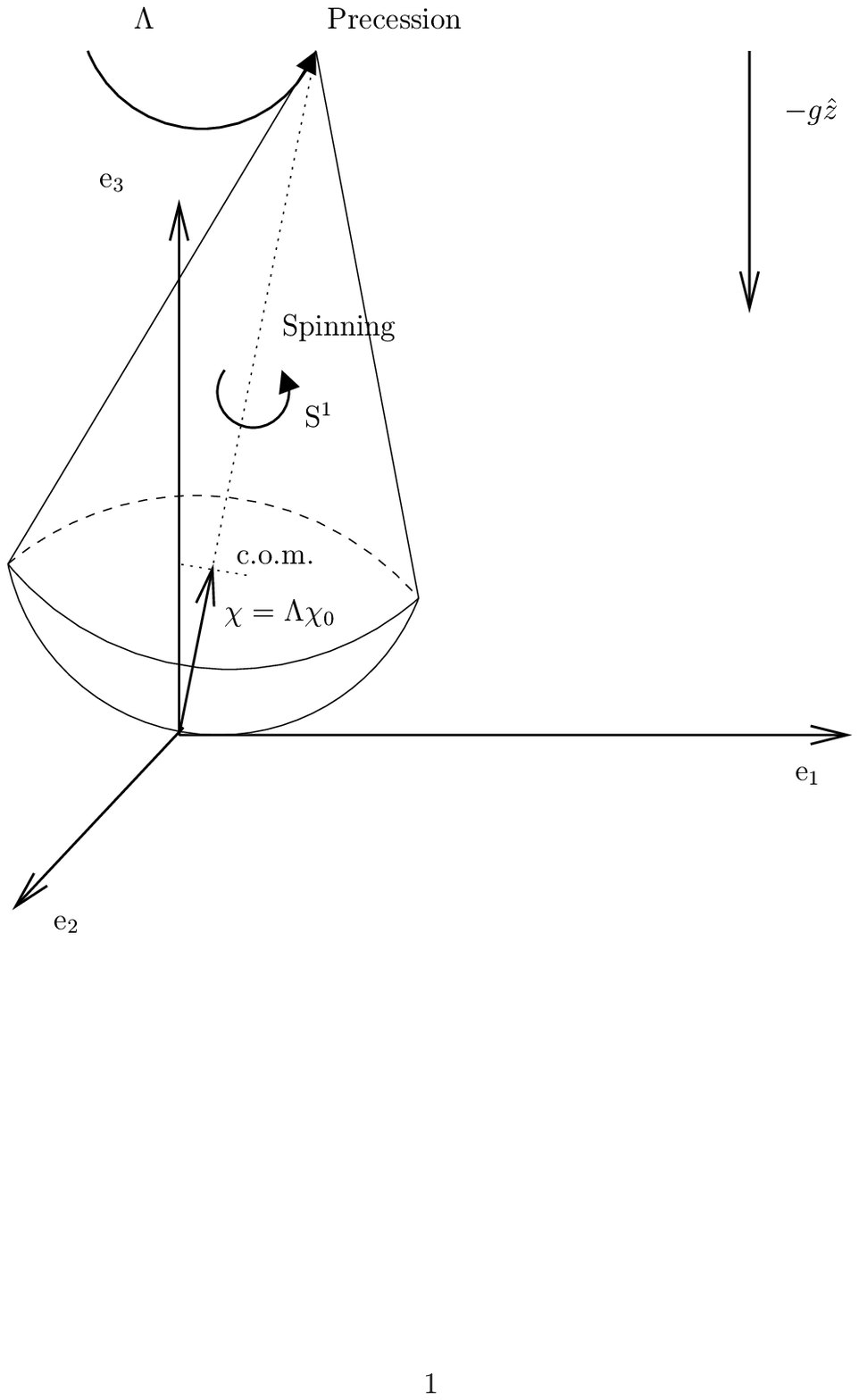}\\
\end{center}
\caption{\scriptsize{The heavy top as viewed in the spatial frame. The heavy body is attached to the spatial frame at an arbitrary point (in this diagram this point is the origin). The motion is composed of two components, precession
and spinning. The vector $\chi$ from the point of support to the centre of
mass of the top (c.o.m.), which lies on the centreline of the top,  rotates about each of the spatial axes with spatial angular velocity $\omega$, maintaining a positive vertical component.
Spatial angular momentum is only conserved for motions purely about $e_3$,
however. The body frame also spins
about its own centreline axis and hence invariance of the Inertia matrix under
the action of $S^1$ is also required for spatial angular momentum conservation.}}
\end{figure}

The Clebsch variable constrained Lagrangian for the discrete heavy top motion in the
spatial representation is

\be
\begin{split}
L&=-\frac{Tr}{h^2}\left(sym(\omega_{k+1})I_{k}\right) - mg \langle\mathbf{\chi}_k,\hat{\mathbf{z}}\rangle   -\frac{Tr}{2}\left(P_{k+1}^T(\Lambda_{k+1}-\omega_{k+1}\Lambda_k)\right)\\
  &-\frac{Tr}{2}\left(J_{k+1}(I_{k+1}-\omega_{k+1}I_k\omega_{k+1}^T)\right)
  -\langle\mathbf{J}^{\chi}_{k+1}, \mathbf{\chi}_{k+1}- \omega_{k+1}\mathbf{\chi}_k\rangle\\
  &+Tr\left(\Theta_{k+1}(\omega_{k+1} \omega_{k+1}^T -I_d)\right).
\end{split}
\ee
The (left) momentum map corresponding to the augmented right diagonal action of $SO(3)$
on $SO(3)\times SO(3)\times V^*\times\mathbb{R}^3$ is

\be
\mathbf{J}^L_{k+1}=\Lambda_k\diamond P_k+
G_k\diamond I_k + \tilde{\mathbf{J}}^{\chi}_{k}\diamond\mathbf{\chi}_{k},
\ee
with  $\tilde{\mathbf{J}}^{\chi}_{k}:=\mathbf{J}^{\chi}_{k}
+mg\hat{\mathbf{z}}$ and $G_k:=\omega_{k+1}^T J_{k+1}\omega_{k+1}$.

The image of $\mathbf{J}^L_{k+1}$ is the spatial angular momentum which is
not conserved unless the motion of the top is about the axis parallel to
the gravity vector. The Clebsch relations give the discrete Euler-Lagrange equations

\be
\begin{split}
\nabla_{\Lambda_{k}}L'(\Lambda_{k-1},\Lambda_{k}) + \nabla_{\Lambda_{k}}L'(\Lambda_{k},\Lambda_{k+1})&=0,\\
\nabla_{I_{k}}L'(I_{k-1},I_{k}) + \nabla_{I_{k}}L'(I_{k},I_{k+1})&=0,\\
\nabla_{\chi_{k}}L'(\mathbf{\chi}_{k-1},\mathbf{\chi}_{k}) + \nabla_{\chi_{k}}L'(\mathbf{\chi}_{k},\mathbf{\chi}_{k+1})&=0.\\
\end{split}
\ee

Evaluation of
the derivatives of $L'$ and subsequent rearrangement gives, respectively,

\be
\begin{split}
P_{k+1}&=\omega_{k+1}P_k,\\
J_{k+1}&=\omega_{k+1}(-(\nabla_{I_k} L)^T + J_k)\omega_{k+1}^T,\\
\mathbf{J}^{\chi}_{k+1}&=\omega_{k+1}\tilde{\mathbf{J}}^{\chi}_k.\\
\end{split}
\ee

Substitution of these three equations into the equation for the (left) momentum
map gives the equations of motion in the spatial variables

\be
\begin{split}
m_{k+1}&=Ad^*_{\omega_{k}^T}m_k - \nabla_{I_k}L\diamond I_k
+mg\mathbf{\chi}_k\diamond \hat{\mathbf{z}},\\
I_{k+1}&=\omega_{k+1}I_k\omega_{k+1}^T,\\
\mathbf{\chi}_{k+1}&=\omega_{k+1}\mathbf{\chi}_k,
\end{split}
\ee
with $m_k:=\frac{2}{h^2}skew\left((\nabla_{\omega_{k+1}}L)^T\omega_{k+1}\right)$.

\newpage
\section{The Coupled Rigid Body}\label{sect:crb}
$\quad$

\begin{figure}[!ht]
\begin{center}
\includegraphics[angle=0.0,width=\textwidth]{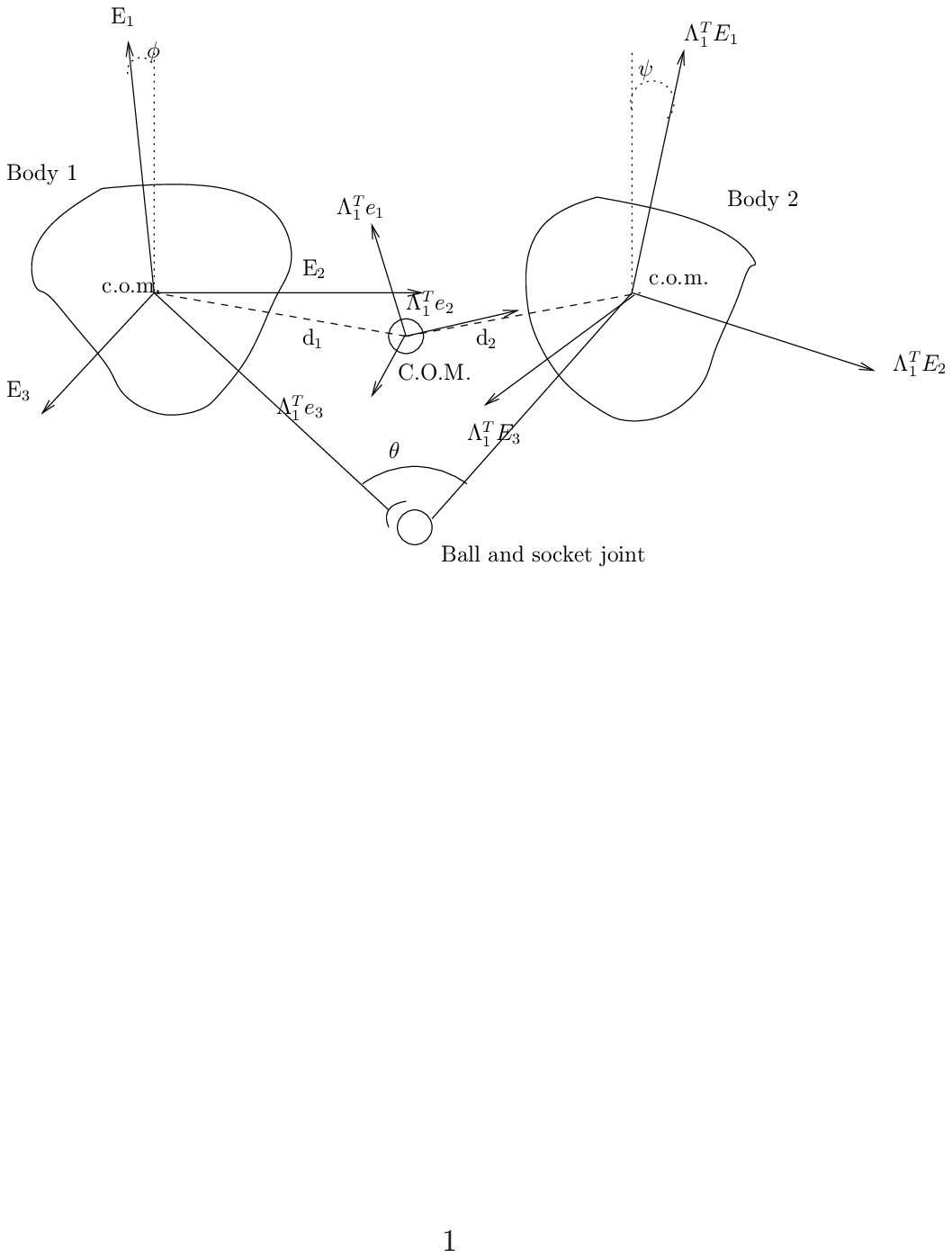}\\
\end{center}
\caption{\scriptsize{The coupled rigid body as viewed in the frame of body 1. Each body is attached from its centre of mass (c.o.m.) to a ball and socket joint. In the $\mathbb{R}^3$
reduced configuration space, the origin of the spatial frame is the centre
of mass (C.O.M.) of the coupled rigid body. In the frame of body 1, the motion is composed of two components, precession
and spinning. The vector $\Lambda d^0_2$, representing the position of the
centre of mass of body 2 in the frame of body 1, rotates about the origin
with relative orientation $\Lambda=\Lambda_1^T\Lambda_2$. $\phi$ and $\psi$
denote the angles between the body axes $E_1$ and the vertical and $\theta$
denotes the angle between the body attachments at the joint. Each body also
spins about its axes, but only the spin of body 2 is observable.}}
\end{figure}

We now state the Clebsch variable constrained action principle, derive the
momentum maps and the equations describing the discrete time free motion of two rigid bodies coupled by a ball and socket joint. We choose the origin of the container to be at the position of center of mass of the coupled body (C.O.M.) (as shown in figure 4) and
let the configuration $\Phi$ be the attitude of body 1 and 2, $\Lambda_1,\Lambda_2\in SO(3)$ relative to a reference configuration. The basic configuration
space, under the assumption that the centre of mass of the coupled body is stationary, is $\mathfrak{C}=SO(3)\times SO(3)$. 

We denote the total mass as $m=m_1+m_2$, the position
of the center of mass of each body as $d_i$ and the attitude of body 2 in the frame of body 1, referred to as the relative orientation matrix, as $\Lambda=\Lambda_1^T\Lambda_2$.

It is also useful to define the
following terms $\epsilon=\frac{m_1m_2}{m}$, $D_{ij}=d_i\otimes d_j$, and
the modified inertia matrix $\hat{I}_i=I_i -\frac{m_i^2}{m}D_{ii}$.
Both $D_{12}$
and $\hat{I}_i$ are fixed in the body frame
whereas $\Lambda$ is an
advected quantity. The discrete Lagrangian on $(SO(3)\times SO(3))^2$ is
\be
L=\sum_{i=1}^2 Tr\left(\Lambda_i^k\hat{I}_i\Lambda_i^{k+1^T}\right) + \epsilon Tr\left((\Lambda_1^{k+1}-\Lambda_1^k)D_{12}(\Lambda_2^{k+1}-\Lambda_2^k)^T\right).
\ee

\subsection{The body representation}

The reduced Lagrangian with Clebsch potentials is
\be
\begin{split}
L'&=\sum_{i=1}^2 Tr\left(\hat{I}_i sym(\Omega_i^{k+1})\right) + \epsilon Tr\left((I_d -\Omega_1^{k+1^T})D_{12}(I_d -\Omega_2^{k+1})\Lambda^{k+1^T}\right)\\
&-\frac{Tr}{2}\left(P_i^{k+1^T}(\Lambda_i^{k+1}-\Lambda_i^k\Omega_i^{k+1})\right)
-\frac{Tr}{2}\left(J^{k+1^T}(\Lambda^{k+1}-\Omega_1^{k+1^T}\Lambda^k\Omega_2^{k+1})\right)\\
&-Tr\left(\Theta_i^{k+1}(\Omega_i^{k+1}\Omega_i^{k+1^T} -I_d)\right).
\end{split}
\ee

The Clebsch relations include the discrete Euler-Lagrange equations

\be
\begin{split}
\nabla_{\Lambda_i^{k}}L'(\Lambda_i^{k-1},\Lambda_i^{k}) + \nabla_{\Lambda_i^{k}}L'(\Lambda_i^{k},\Lambda_i^{k+1})&=0,~~i\in\{1,2\},\\
\nabla_{\Lambda^{k}}L'(\Lambda^{k-1},\Lambda^{k}) + \nabla_{\Lambda^{k}}L'(\Lambda^{k},\Lambda^{k+1})&=0.\\
\end{split}
\ee
Evaluation of
the derivatives of $L'$ and subsequent rearrangement gives, respectively,

\be \label{eqn:flowAugCoTB_DT}
\begin{split}
P_i^{k+1}&=P_i^k\Omega_i^{k+1},\\
J^{k+1}&=\Omega_1^{k+1^T} J^k\Omega_2^{k+1}.\\
\end{split}
\ee

Addition of the Clebsch relations paired with $\delta \Omega_i^{k+1}$ gives the expression

\be
J^R=\sum_{i=1}^2 \text{skew}(\Lambda_i^{k^T} P_i^k) + [J^k,\Lambda^k], 
\ee
which satisfies the definition of the momentum map for cotangent lifted left
actions
of SO(3)

\be
\begin{split}
\langle J^R, \zeta \rangle&= \sum_{i=1}^2\langle P^k_i, \mathfrak{L}_{\zeta}\Lambda^k_i\rangle
+\langle J^k, \mathfrak{L}_{\zeta}\Lambda^k\rangle\\
&=\langle \sum_{i=1}^2 P^k_i\diamond \Lambda^k_i+ J^k\diamond\Lambda^k,\zeta\rangle.
\end{split}
\ee

\begin{lem}
The total spatial angular momentum is conserved by the discrete time flow on the augmented bundle.
\end{lem}

\begin{proof}
The total spatial angular momentum is
\be
\begin{split}
m^{k+1}=m_1^{k+1} + m_2^{k+1}&=\sum_{i=1}^2\Lambda_i^{k}M_i^{k+1}\Lambda_i^{k^T}\\
&=\sum_{i=1}^2 2skew(\Lambda_i^{k}P_i^{k^T} + \Lambda_1^kJ^k\Lambda_2^{k^T}
+\Lambda_2^k J^{k^T}\Lambda_1^{k^T})\\
&=\sum_{i=1}^2 2skew(\Lambda_i^{k-1}\Omega_i^{k}\Omega_i^{k^T} P_i^{{k-1}^T})\\
&=m_1^k+m_2^k=m^k.
\end{split}
\ee
\end{proof}

Substituting the expressions in equation \ref{eqn:flowAugCoTB_DT} into the
right momentum map gives the discrete EP equations in body
variables

\be
M_i^{k+1}=Ad^*_{\Omega_i^{k^T}} M_i^k, ~~i\in\{1,2\}\\
\ee
together with the evolution equation for the relative orientation matrix
in the frame of body $1$
\be
\Lambda^{k+1}=\Omega_1^{k+1^T}\Lambda^k\Omega_2^{k+1}.
\ee

\subsection{The spatial representation}
The Clebsch variable constrained Lagrangian for the spatial representation
is
\be
\begin{split}
L'&=\sum_{i=1}^2 Tr\left(\hat{I}^k_i sym(\omega_i^{k+1})\right) + \epsilon Tr\left((\omega_1^{k+1}-I_d)D^k_{12}(\omega_2^{k+1^T}-I_d)\right)\\
&-Tr\left(P_i^{k+1^T}(\Lambda_i^{k+1}-\omega_i^{k+1}\Lambda_i^k)\right)
-Tr\left(J_i^{k+1^T}(\hat{I}_i^{k+1}-\omega_i^{k+1}\hat{I}^k_i\omega_i^{k+1^T})\right)\\
&-Tr\left(J^{k+1^T}(D_{12}^{k+1}-\omega_1^{k+1}D_{12}^k\omega_2^{k+1^T})\right)
-Tr\left(\Theta_i^{k+1}(\omega_i^{k+1^T}\omega_i^{k+1} -I_d)\right).
\end{split}
\ee
where $\hat{I}^k_i=\Lambda^k_i\hat{I}_i\Lambda_i^{k^T}$ and $D^k_{12}=\Lambda^k_1D_{12}\Lambda_2^{k^T}$
are now time dependent.

The Clebsch relations include the discrete Euler-Lagrange equations
\be
\begin{split}
\nabla_{\Lambda_i^{k}}L'(\Lambda_i^{k-1},\Lambda_i^{k}) + \nabla_{\Lambda_i^{k}}L'(\Lambda_i^{k},\Lambda_i^{k+1})&=0,~~i\in\{1,2\},\\
\nabla_{\hat{I}_i^{k}}L'(\hat{I}_i^{k-1},\hat{I}_i^{k}) + \nabla_{\hat{I}_i^{k}}L'(\hat{I}_i^{k},\hat{I}_i^{k+1})&=0,~~i\in\{1,2\},\\
\nabla_{D_{12}^{k}}L'(D_{12}^{k-1},D_{12}^{k}) + \nabla_{D_{12}^{k}}L'(D_{12}^{k},D_{12}^{k+1})&=0.\\
\end{split}
\ee

Evaluation of
the derivatives of $L'$ and subsequent rearrangement gives, respectively,

\be \label{eqn:flowAugCoTB_DTS}
\begin{split}
P_i^{k+1}&=\omega_i^{k+1}P_i^k,\\
J_i^{k+1}&=\omega_i^{k+1}\underbrace{\left(-sym(\omega_i^k) + J_i^k\right)}_{G_i^k}\omega_i^{k+1^T},\\
J^{k+1}&=\omega_1^{k+1}J^k\omega_2^{k+1^T}.
\end{split}
\ee

Addition of the Clebsch relations paired with $\delta \Omega_i^{k+1}$ gives the expression

\be
J^L=\sum_{i=1}^2 skew(P^k_i\Lambda_i^{k^T})  +[G^k_i,\hat{I}^k_i] + \text{skew}([J^k, D_{21}^{k}]),
\ee
which satisfies the definition of the left momentum map for a cotangent lifted right action of $SO(3)$

\be
\begin{split}
\langle J^L,\zeta\rangle&=\sum_{i=1}^2\langle P^k_i, \mathfrak{L}_{\zeta}\Lambda^k_i \rangle + \langle G^k_i, \mathfrak{L}_{\zeta}\hat{I}^k_i\rangle +  \langle J^k, \mathfrak{L}_{\zeta}D^k_{21}\rangle\\
&=\langle \sum_{i=1}^2 P^k_i\diamond \Lambda^k_i + G^k_i\diamond \hat{I}^k_i  +
J^k\diamond D_{21}^k,\zeta\rangle.
\end{split}
\ee

The total spatial angular momentum $m=m_1+m_2$ is the image of $J^L$ where
the spatial angular momentum of each body is
\be
\begin{split}
m^k_1&=2skew(\hat{I}_1^k\omega_1^{k+1} -\epsilon \underbrace{\omega_1^{k+1^T}(\omega_2^{k+1}-I_d)D_{12}^{k^T}}_{\Gamma_1^{k+1}}),\\
m^k_2&=2skew(\hat{I}_2^k\omega_2^{k+1} -\epsilon\underbrace{\omega_2^{k+1^T}(\omega_1^{k+1}-I_d)D_{12}^{k}}_{\Gamma_2^{k+1}}).
\end{split}
\ee

In summary, the discrete EP equations of motion in the spatial frame are

\be
\begin{split}
m^{k+1}&=\sum_{i=1}^2 Ad^*_{\omega^k_i}m^{k+1}_i + \nabla_{\hat{I}^k_i}L\diamond\hat{I}^k_i,\\
\hat{I}_i^{k+1}&=\omega_i^{k+1}\hat{I}^k_i\omega_i^{k+1^T},\\
D_{12}^{k+1}&=\omega_1^{k+1}D_{12}^k\omega_2^{k+1^T}.\\
\end{split}
\ee
We remark that these equations take a very similar form for the $n$ multi-body
in which the summation is over $1\rightarrow n$. Tables 5 and 6 of appendix A summarise the main results of this section. We
present results from the implementation of the body representation of the
discrete EP equations with advected parameters in the next section.

\section{Numerical Experiments}\label{sect:num_exp}

This section presents results demonstrating the conservative properties and
accuracy of the rigid body, heavy top and coupled rigid body integrators.
The components of the body momentum are compared with the analytic solution
for the rigid body only, and the Matlab Ode45 integration of the Euler-Arnold
ordinary differential equations and their variants for the heavy top and coupled rigid body. 
The tolerance of the Ode45 routine is set to 1e-15. The time step for all numerical experiments
is $\Delta t=0.1$. Although the figure captions give details of each experiment, we point out
a few general features here.

\vspace{1cm}
\begin{itemize}
\item  Firstly, the choice of initial parameters in each experiment avoids intersection of the body momenta with fixed
points. It is well known that the coadjoint orbits of the classical rigid body with distinct moments of inertia have saddle points at
$(0,\pm\pi, 0)$ (which are connected by four heteroclinic orbits) and centers at $(\pm\pi, 0, 0)$ and $(0, 0,\pm\pi)$. The numerical solution does not
become unstable, however, provided the time step is no larger
than approximately 0.5.  
\item The numerical round-off error in each representation depends upon the principle moments of inertia. This is shown after $10^4$ time steps in figures 7 and 8. For example,
when $I_1=I_2>I_3$, the error in the (i) spatial angular momentum is $O(10^{-8})$
and $O(10^{-11})$ and (ii) energy is $O(10^{-7})$
and $O(10^{-10})$ for the respective body and spatial representations. When
$I_1>I_2>I_3$, the same respective errors in the (i) spatial angular momentum are $O(10^{-11})$
and $O(10^{-14})$ and (ii) energy are $O(10^{-10})$
and $O(10^{-13})$.
\item The results comparing Dormand and Prince's explicit Runge-Kutta method
\cite{DORMAND80} (implemented in the
ode45 routine) and the DMV algorithms should not be interpreted as a performance
comparison. In each experiment, the ode45 solver was run at the smallest
time step possible purely to provide a quantitative benchmark for the DMV
algorithm.
\item We find a good agreement between the numerical results
and the analytic solution for the rigid body and confirm conservation of
spatial angular
momentum and the Casimirs $||M||_2^2$ and $(det(I), || I||)$ for the body
and spatial representations respectively. For the body representation of heavy top motion, the Casimir
$\langle M,\Gamma\rangle$ is also correctly computed. Our numerical results qualitatively match those obtained
by McLachlan and Zanna \cite{MCLACHLAN05} and Celledoni
and S\"afstr\"om \cite{CELLEDONI05} for the body representation of
the rigid body and heavy top respectively.
\item We perform two coupled rigid body experiments in which two identical bodies are subject to the same initial
conditions (i) but are initially at right angles to each other and (ii)
are initially aligned with each other. In the first experiment, shown in figure
10 we observe
non-periodic behaviour in the body angular momentum components caused by
exchanges of momentum between the two bodies. The second component of the
momentum changes the most, ranging from $-1$ to $0.8$. In the second experiment, shown in figure
11, we recover a rigid
body motion similar to that shown in figure 5, except that $M_3$ varies. 
\end{itemize}

\begin{figure}[!ht] 
\includegraphics[angle=0.0,width=0.9\textwidth]{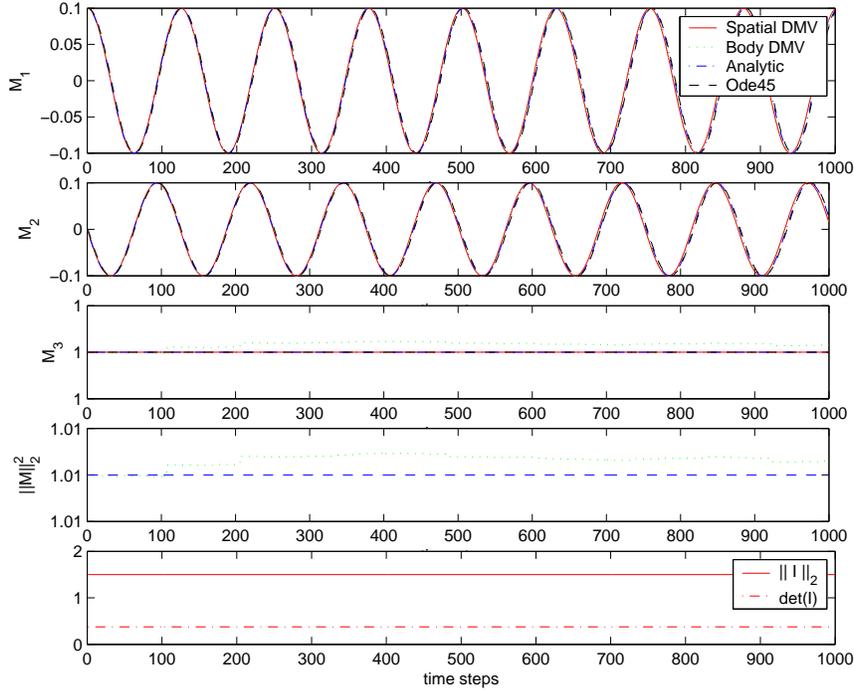}\\
\caption{This figure compares numerical simulations of a rigid body over
$1000$ time steps for the case when the principal moments of inertia $I_1=I_2>I_3$ ($I_1=2,~I_2=2,~I_3=1$). The top three graphs each show a component of the body angular momentum 
of rigid body motion. The bottom two graphs show the Casimirs $||M||_2^2$ and $||I||_2,~det(I)$
of rigid body flow in the respective body and spatial representations. The
graph labelled (i) 'body DMV' is the solution computed by the body DMV integrator, (ii) 'spatial DMV' is the body frame translated solution computed by the spatial DMV integrator (which computes the angular momentum and moment of inertia in the spatial representation) (iii) 'ode45' is an explicit Runge-Kutta (4,5) integrated solution of the Euler-Arnold equations using the Matlab routine,
ode45 and (iv) 'analytic' is the analytical solution. The initial conditions
for this simulation are the initial body angular momentum components given
as $M_1(0)=0.1,~M_2(0)=0,~M_3(0)=1$. The top three graphs show that the DMV momentum matches the analytical solution which describes the rolling of a cone of constant angle in the body on a second cone of constant angle fixed in space \cite{MARSDEN99a}. The $2^{nd}$ from bottom graph shows that the body DMV integrator precisely computes the Casimir $||M||_2^2$ \cite{MARSDEN99a} suggesting preservation of the rigid body Lie-Poisson structure
and consequently that the DMV angular momentum remains on the sphere. The
bottom graph shows that the Casimirs $||I||_2$ and $det(I)$ of the spatial DMV integrator
are correctly computed suggesting that the Lie-Poisson structure on the dual of the
semi-direct product lie algebra is also preserved.}
\end{figure}


\begin{figure}[!ht] 
\includegraphics[angle=0.0,width=0.9\textwidth]{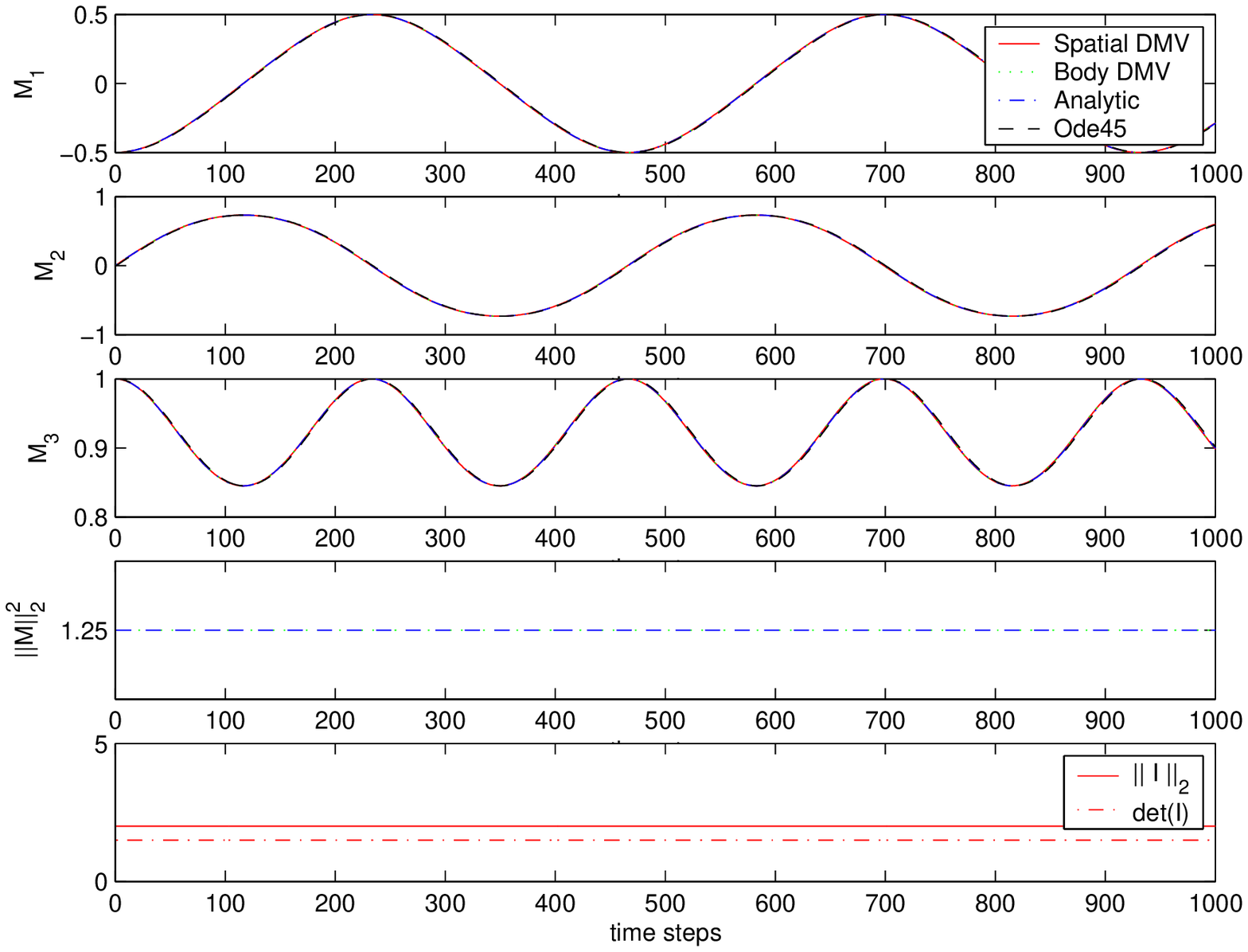}\\
\caption{This figure compares numerical simulations of a rigid body over
$1000$ time steps for the case when the principal moments of inertia $I_1>I_2>I_3~(I_1=3.5,~I_2=2.5,~I_3=2)$. The top three graphs each show a component of the body angular momentum 
of rigid body motion and the bottom two show the Casimirs $||M||_2^2$ and $||I||_2,~det(I)$
of rigid body flow in the respective body and spatial representations. 
The initial conditions
for this simulation are the initial body angular momentum components given
as $M_1(0)=-0.5,~M_2(0)=0,~M_3(0)=1$. The top three graphs show that the DMV momentum matches the analytical solution describing the intersection of energy ellipsoids with coAdjoint orbits of $SO(3)$ which are two-spheres \cite{MARSDEN99a}. Note that although
our choice of simulation parameters avoids the flow intersecting either of the two saddle points at $(0,\pm ||M||_2, 0)$ or centers at $(\pm ||M||_2,0,0)$ and $(0,0,\pm ||M||_2)$, the solution does not become unstable provided the
maximum time step is $\approx 0.5$.}
\end{figure}

\begin{figure}[!ht] 
\includegraphics[angle=0.0,width=0.9\textwidth]{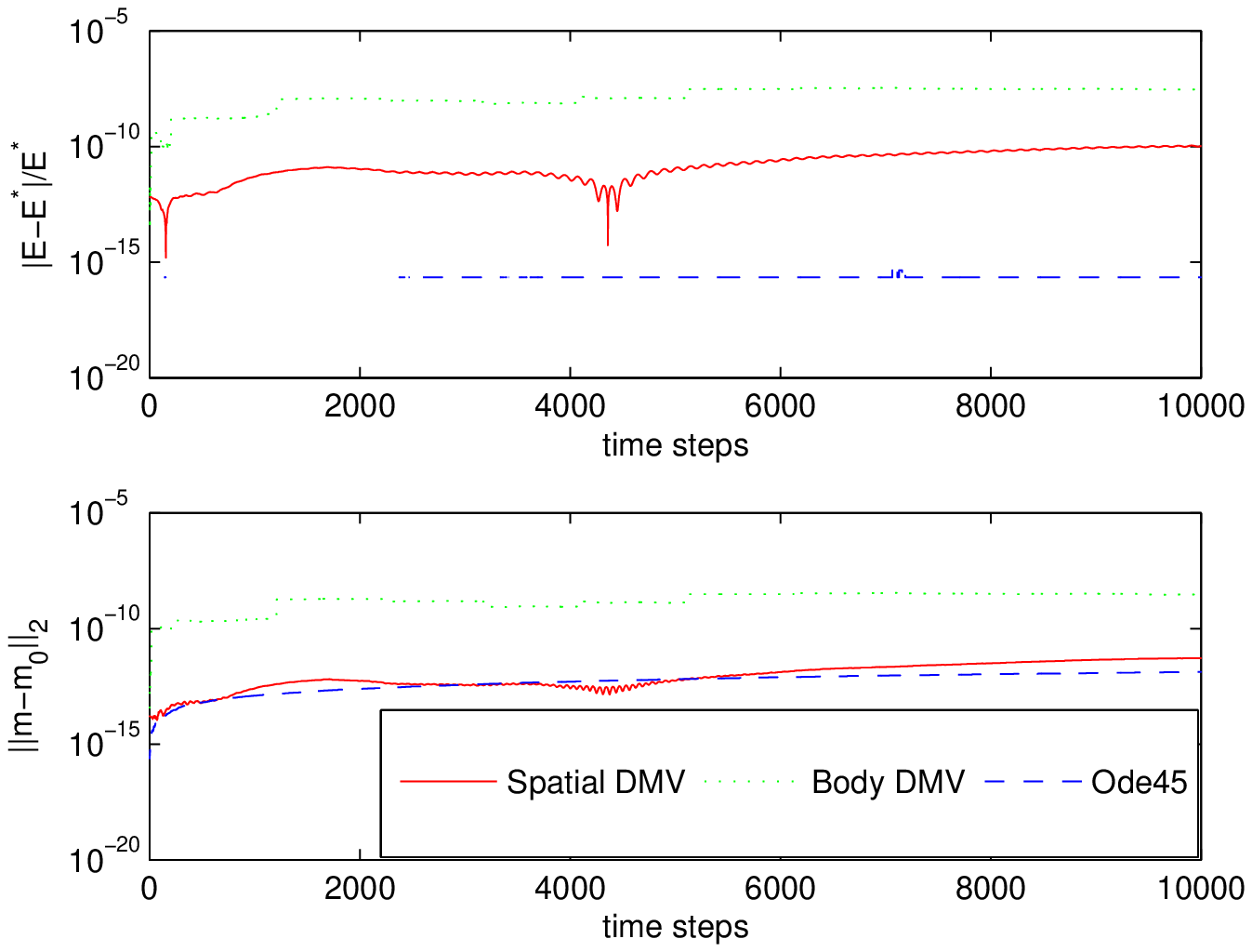}\\
\caption{This figure compares the energy and spatial angular
momentum error in numerical simulations of rigid body motion over $10000$ time
steps for the case when the principal moments of inertia $I_1=I_2>I_3$
$(I_1=2,~I_2=2,~I_3=1)$. The top graph shows the relative energy error growth in the solutions of
the 'DMV' and 'ode45' integrators and the bottom graph shows the error in the approximated spatial angular momentum. The initial conditions
for this simulation are the initial body angular momentum components given
as $M_1(0)=0.1,~M_2(0)=0,~M_3(0)=1$.
The graphs show that the error in the body DMV integrator computation of
the energy and spatial angular momentum is higher than the error computed
by the spatial DMV integrator. The bottom graph also shows that the spatial DMV integrator
conserves spatial angular momentum to numerical round off.}
\end{figure}

\begin{figure}[!ht] 
\includegraphics[angle=0.0,width=0.9\textwidth]{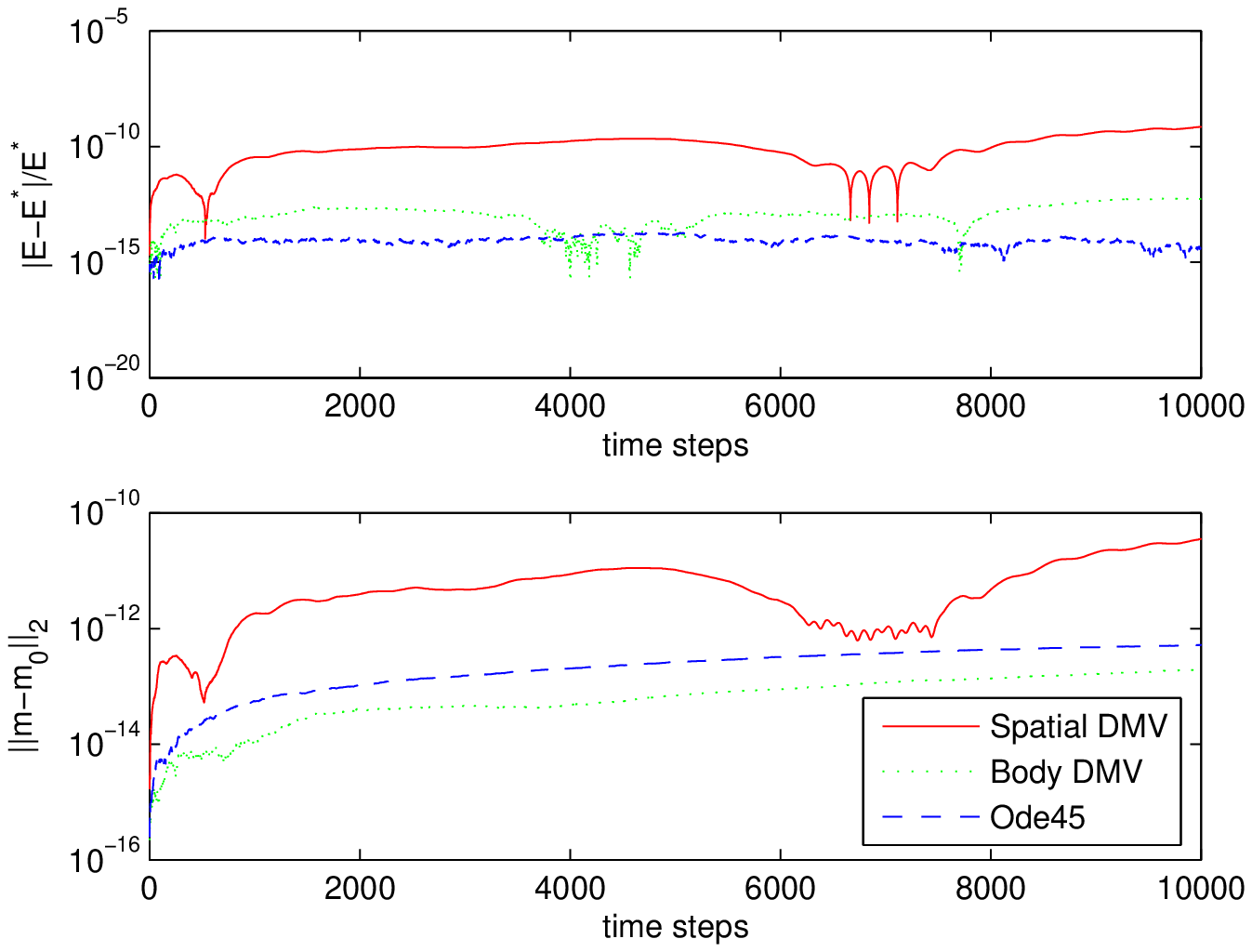}\\
\caption{This figure compares the energy and spatial angular
momentum error in numerical simulations of rigid body motion over $10000$ time steps for the case when the principal moments of inertia $I_1>I_2>I_3$
$(I_1=3.5,~I_2=2.5,~I_3=2)$. The top graph shows the relative energy error growth in the solutions of
the 'DMV' and 'ode45' integrators and the bottom graph shows the error in the approximated spatial angular momentum. The initial conditions
for this simulation are the initial body angular momentum components given
as $M_1(0)=-0.5,~M_2(0)=0,~M_3(0)=1$.
In contrast with the previous figure, the graphs show that the error in the spatial DMV integrator computation of the energy and spatial angular momentum is higher than the error computed by the body DMV integrator. The bottom graph also shows that the body DMV integrator conserves spatial angular momentum to numerical round off.}
\end{figure}

\begin{figure}[!ht] 
\includegraphics[angle=0.0,width=0.9\textwidth]{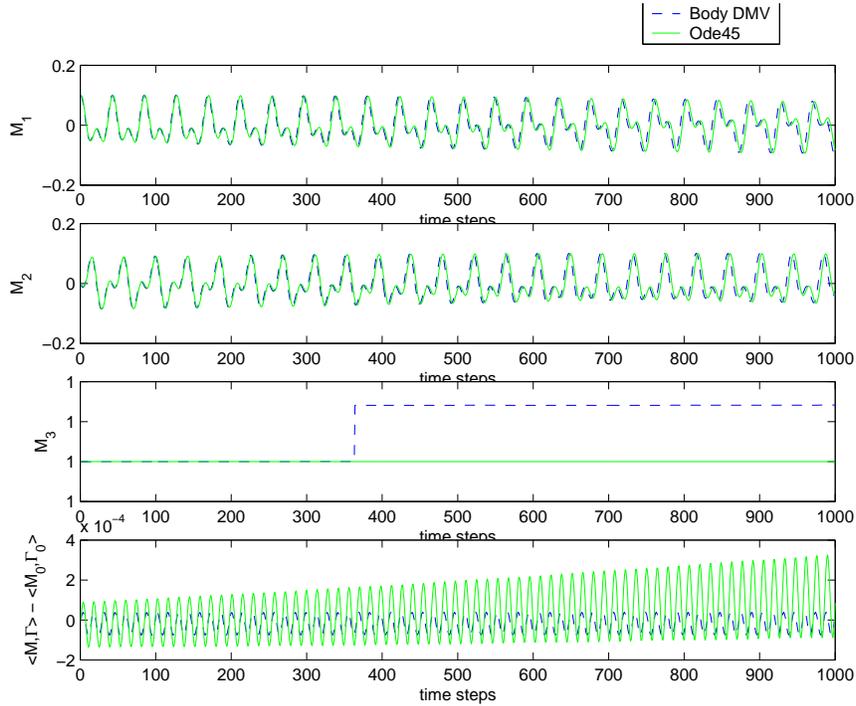}\\
\caption{This figure compares numerical simulations of the body representation
of the heavy top over
$1000$ time steps for the case when the principal moments of inertia $I_1=I_2>I_3$ ($I_1=2,~I_2=2,~I_3=1$). The top three graphs each show a component of the body angular momentum 
of heavy top motion and the bottom graph shows the error in the 'body DMV'
and 'ode45' computed Casimir $\langle M,\Gamma\rangle$ of the heavy top Lie-Poisson bracket. The initial conditions
for this simulation are the initial (i) body angular momentum components and (ii) position of the vertical axis in the body frame
given respectively as $M_1(0)=0.1,~M_2(0)=0,~M_3(0)=1$ and $\Gamma=[0,0,1]$.
Whenever the first and second components of the body angular momentum are non-zero, heavy top motion breaks the $S^1$ symmetry about the vertical axis.
The bottom figure confirms that the Casimir $\langle M,\Gamma\rangle$ is always conserved.}
\end{figure}

\begin{figure}[!ht] 
\includegraphics[angle=0.0,width=0.9\textwidth]{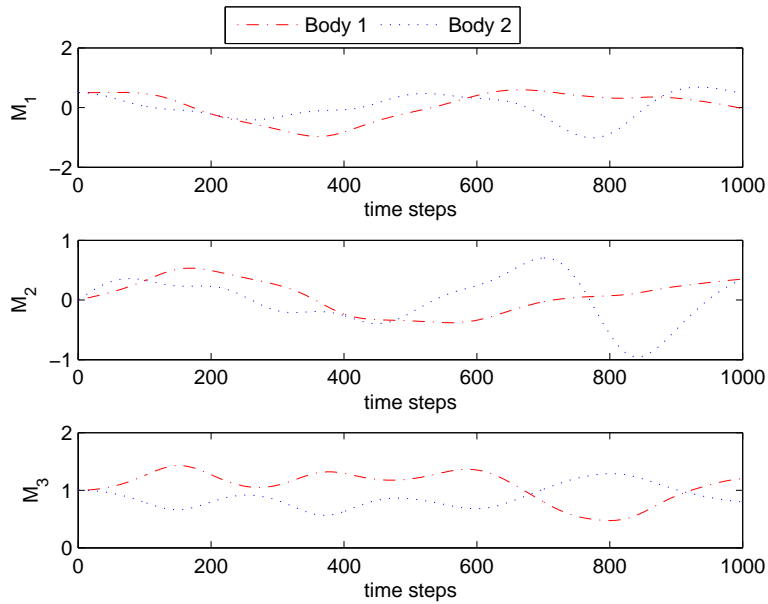}\\
\caption{This figure compares numerical simulations of the coupled rigid
body, as seen in the frame of body 1, over
$1000$ time steps for the case when the two identical rigid bodies are initially
positioned at right angles to each other. The top three graphs each show a component of the body angular momentum 
of body 1 and 2. The initial conditions
for this simulation are the initial (i) body angular momentum components
(ii) orientation of the bodies relative to their $E_3$ axes and (iii)
angle between the mechanical attachments at the ball and socket joint
given respectively as $M^2(0)=M^1(0)=[0.5,0,1]$, $\phi(0)=\psi(0)$ and
$\theta(0)=\frac{\pi}{2}$. The principal moments of inertia of the
two identical rigid bodies are $I_1=I_2>I_3$ ($I_1=2,~I_2=2,~I_3=1$). 
The graphs show that the components of body angular momentum are not periodic
and extreme values are different from those of the single (uncoupled) rigid body shown in figure 5, indicating
transfer of angular momentum between the two bodies.}
\end{figure}
\begin{figure}[!ht] 
\includegraphics[angle=0.0,width=0.9\textwidth]{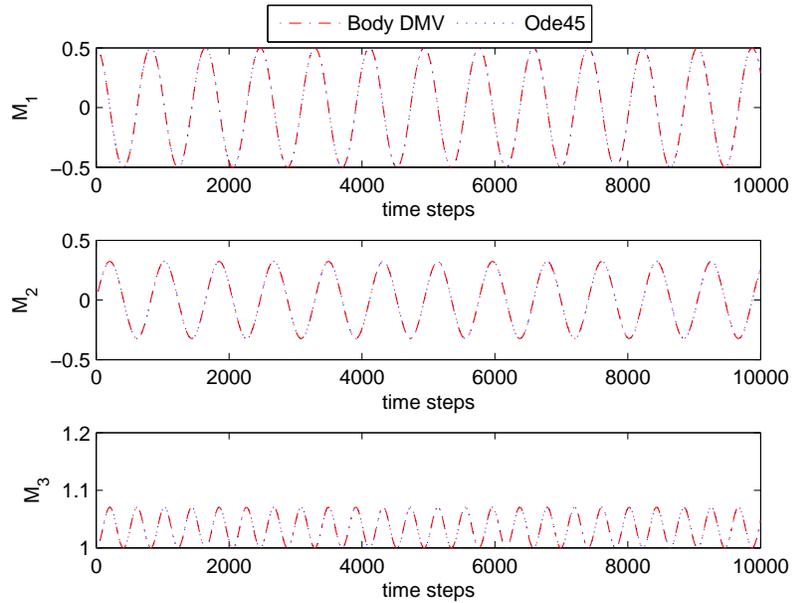}\\
\caption{This figure compares numerical simulations of the coupled rigid
body, as seen in the frame of body 1, over
$1000$ time steps for the case when the two identical rigid bodies are initially
aligned with each other. The top three graphs each show a component of the body angular momentum 
of body 1 and 2. The initial conditions
for this simulation are the initial (i) body angular momentum components
(ii) orientation of the bodies relative to their $E_3$ axes and (iii)
angle between the mechanical attachments at the ball and socket joint
given respectively as $M^2(0)=M^1(0)=[0.5,0,1]$, $\phi(0)=\psi(0)$ and
$\theta(0)=0$. The principal moments of inertia of the
two identical rigid bodies are $I_1=I_2>I_3$ ($I_1=2,~I_2=2,~I_3=1$). 
The graphs show that the components of body angular momentum of both bodies
are similar to those of the rigid body motion shown in figure 5, differing in the $M_3$ component, which varies here.}
\end{figure}

\begin{figure}[!ht]
\includegraphics[angle=0.0,width=0.9\textwidth]{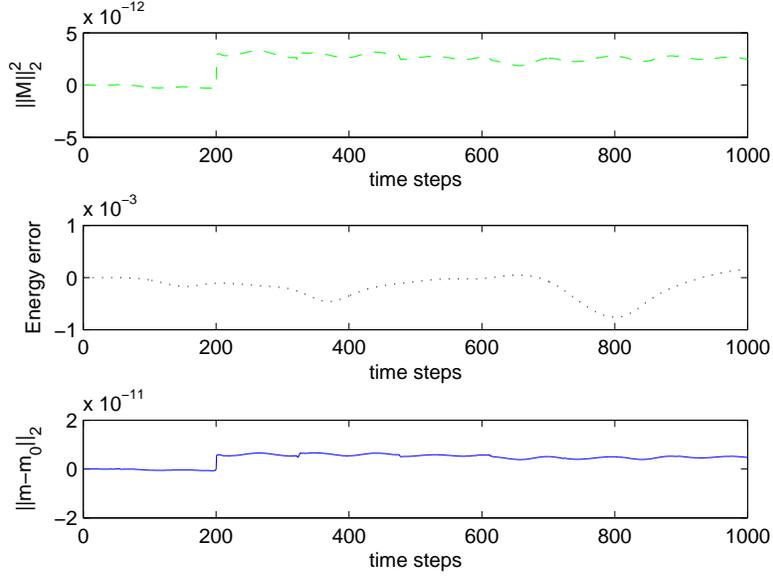}\\
\caption{This figure shows the error in computation of the Casimirs and conserved
spatial angular momentum of the coupled rigid body motion, as viewed in the frame
of body 1, for the case when both
identical bodies are initially positioned at right angles to each other. The top graph shows the absolute error in 'DMV' and 'ode45' computation of the C.R.B. Casimir $||M||_2^2=||M_1 + \Lambda M_2 \Lambda^T||_2^2$
\cite{GROSSMAN88}. The middle and bottom graphs show the comparative error
in the energy and spatial angular momentum of the coupled rigid body. The initial conditions
for this simulation are the initial (i) body angular momentum components (ii) orientation of the bodies relative to their $E_3$ axes and (iii)
angle between the mechanical attachments at the ball and socket joint
given respectively as $M^2(0)=M^1(0)=[0.5,0,1]$, $\phi(0)=\psi(0)$ and
$\theta(0)=\frac{\pi}{2}$. The principal moments of inertia of the
two identical rigid bodies are $I_1=I_2>I_3$ ($I_1=2,~I_2=2,~I_3=1$).}
\end{figure}

\newpage
\clearpage
\section{Conclusion}

The spatial and body representations of rigid body motion correspond to the spatial and convective representations of continuum dynamics. The discrete
Clebsch approach \cite{COTTER06} provides a unified framework to derive variational integrators for both descriptions of the continuum. In this paper we demonstrate
the utility of this framework by deriving explicit variational integrators for various rigid body motions in
the spatial and body representations. We derive the momentum maps corresponding
to the symmetry reductions of the discrete Lagrangians, the discrete EP
equations and consequently prove, where appropriate, conservation of spatial angular momentum. 

This paper has pursued the
relationship between variational integrators, derived in the discrete Clebsch framework, and existing studies of discrete integrable rigid body systems. In the body representation, we recover a discrete integrable
analogue of the Euler-Arnold equations, first discovered by Moser and Veselov
\cite{MOSER91}. In the spatial representation we obtain discrete EP
equations with an advected parameter. These correspond to the Lie-Poisson equations on the dual space of a semidirect product lie algebra discovered by Bobenko and
Suris \cite{BOBENKO99}. This discovery provides a discrete extension to the work by
Holm, Marsden and Ratiu \cite{HOLM98} who developed the theory of EP
entirely within a Lagrangian framework so that the EP equations with advection
always correspond to Lie-Poisson Hamiltonian systems
on the dual of a semi-direct
product Lie-algebra. Consequently,
the DMV equations have a family
of Casimirs associated with the Lie-Poisson bracket for these systems, two of which we confirmed by numerical simulation of the spatial
representation of the rigid
body.

We provide several numerical experiments to demonstrate
the conservative properties and accuracy of the explicit variational integrators that we derived
in each case. 
We find a good agreement between the numerical results
and the analytic solution for the rigid body and detect conservation of the
spatial angular
momentum and the total body angular momentum Casimir when the
heavy top motion precesses purely about the vertical axis. Additionally, the numerical results in this paper qualitatively match those obtained
by McLachlan and Zanna \cite{MCLACHLAN05} for the rigid body and Celledoni
and S\"afstr\"om \cite{CELLEDONI05} for the heavy top.  

We also observe
non-periodic behaviour and a four-fold increase in the extremal values of the body angular momentum (relative to the uncoupled case) caused by
exchanges of momentum between two identical coupled rigid bodies, which are
both initialised
with the same body angular momentum, but initially positioned at right angles to each other. 

We do, however, observe some less favourable and unexpected features
in the numerical experiments. Firstly, in the rigid body experiments, the numerical solution becomes unstable
and unconservative if the coAdjoint orbits on the sphere intersect either of the
saddle points at
$(0,\pm\pi, 0)$ (which are connected by four heteroclinic orbits) or centers at $(\pm\pi, 0, 0)$ and $(0, 0,\pm\pi)$ unless the time step is approximately
no larger than $0.5$. Precise bounds on the time step should be determined
before applying DMV integrator to the study of bifurcating systems
such as the heavy top by Lewis, Simo and Marsden \cite{LEWIS92} and the dynamics
and stability of coupled rigid bodies by Sreenath, Krishnaprasad and Marsden
\cite{SREENATH88} and geometrically exact rod by Simo, Posbergh and Marsden \cite{SIMO90}. 

Secondly, we find that the relative
round-off error growth between the two representations is dependent upon
the principle moments of inertia. For example, when $I_1=I_2>I_3$, the error in the (i) spatial angular momentum is $O(10^{-8})$
and $O(10^{-11})$ and (ii) energy is $O(10^{-7})$
and $O(10^{-10})$ for the respective body and spatial representations. When
$I_1>I_2>I_3$, the same respective errors in the (i) spatial angular momentum are $O(10^{-11})$
and $O(10^{-14})$ and (ii) energy are $O(10^{-10})$
and $O(10^{-13})$. For very long simulations, this aspect
should not be overlooked and there may be practical
benefit in choosing one representation over the other for the purpose of
minimising round-off error.

The discrete and continuous versions of the equations of motion and momentum maps are remarkably similar. Appendix A compares the body and spatial representations of these quantities in their continuous and discrete forms.

In a forthcoming paper, we will attempt to address outstanding numerical issues and describe how the discrete
Clebsch approach can be extended for ellipsoidal motions and motions of a geometrically
exact rod in the convective representation.

\newpage
\appendix
\section{Body and Spatial Representations in Continuous
and Discrete Time}

\begin{small}
\begin{table}[!ht]
\begin{tabular}{lllllll}
\hline
\vline & Property &\vline  & Continuous &\vline  & Discrete &\vline \\
\hline
\vline & Body attitude  & \vline  & $\Lambda(t)\in SO(N)$  & \vline  &  $\Lambda_k\in SO(N)$ & \vline \\ \vline & Angular velocity  & \vline  &  $\Omega=\Lambda^T\dot{\Lambda}=-\Omega^T$  & \vline  & $\Omega_{k+1}=\Lambda_{k}^T\Lambda_{k+1}$ & \vline \\ 
\vline & Inertia Matrix  & \vline  & $I_0$ & \vline  &  & \vline \\ \vline &
Angular momentum  & \vline  & $M= I_0\Omega -\Omega^T I_0$  & \vline  & $M_k=I_{0}\Omega_k - \Omega_k^T I_0$ & \vline \\ \vline &
Equations of motion  & \vline  & $\dot{M}=ad^*_{\Omega}M$  & \vline  & $M_{k+1}=Ad^*_{\Omega_k^T}M_k$ & \vline \\ \vline &
Right momentum map  & \vline  & $J^R=P\diamond \Lambda$ & \vline  & $J_R^{k+1}=P_{k}\diamond\Lambda_{k}$ & \vline \\ 
\hline\\
\end{tabular}
\caption{Comparison of the terms required to describe the \emph{body} representation of the \textbf{rigid body} in continuous and discrete time as derived using
the Clebsch approach. Blank items in the right-hand
column indicate that they are identical to their discrete time descriptions.}
\end{table}

\begin{table}[!ht]
\begin{tabular}{lllllll}
\hline
\vline & Property &\vline & Continuous &\vline & Discrete & \vline\\
\hline
\vline & Body attitude  & \vline  & $\Lambda(t)\in SO(N)$ & \vline  & $\Lambda_k\in SO(N)$ & \vline \\ \vline &
Angular velocity  & \vline  &  $\omega=\dot{\Lambda}\Lambda^T=-\omega^T$  & \vline  &$\omega_{k+1}=\Lambda_{k+1}\Lambda_{k}^T$ & \vline \\ \vline &
Inertia Matrix  & \vline  & $I=\Lambda I_0 \Lambda^T$  & \vline  & $I_k=\Lambda_k I_0 \Lambda_k^T$ & \vline \\ \vline &
Angular momentum  & \vline  & $m=I\omega- \omega^T I$  & \vline  & $m_k=I_{k}\omega_k - \omega_k^T I_k$ & \vline \\ 
\hline
\vline &
Equations of motion  & \vline  & $\dot{m}=ad^*_{\omega}m - \nabla_I L\diamond I=0$,  & \vline  &
$m_{k+1}=Ad^*_{\omega_k^T}m_k - \nabla_{I_k}L\diamond I_k$, & \vline \\ \vline &
 & \vline  & $\dot{I}=[\omega,I]$  & \vline  & $I_{k+1}=\omega_{k+1}I_k\omega_{k+1}^T$  & \vline \\ 
 \hline
 \vline &
Left momentum map  & \vline  & $J^L=P\diamond \Lambda + J\diamond I$ & \vline  & $J_L^{k+1}=P_{k}\diamond\Lambda_{k} + G_k\diamond I_k$ & \vline \\ 
\hline\\
\end{tabular}
\caption{Comparison of the terms required to describe the \emph{spatial} representation of the \textbf{rigid body} in continuous and discrete time.}
\end{table}

\newpage

\begin{table}[!ht]
\begin{tabular}{lllllll}
\hline
\vline&Property  & \vline& Continuous  & \vline& Discrete & \vline\\
\hline
\vline&Body attitude  & \vline& $\Lambda(t)\in SO(N)$ & \vline& $\Lambda_k\in SO(N)$  & \vline\\
\vline& Angular velocity  & \vline&  $\Omega=\Lambda^T\dot{\Lambda}=-\Omega^T$  & \vline& $\Omega_{k+1}=\Lambda_{k}^T\Lambda_{k+1}$ & \vline\\
\vline& Inertia Matrix  & \vline& $I_0$  & \vline& & \vline\\
\vline&Angular momentum  & \vline& $M=I_0\Omega + \Omega^T I_0$  & \vline& $M_k=I_{0}\Omega_k - \Omega_k^T I_0$ & \vline\\
\vline& Orientation of the z-axis  & \vline& $\mathbf{\Gamma}=\Lambda^T\mathbf{z}$  & \vline& $\mathbf{\Gamma}_k=\Lambda_k^T\mathbf{z}$ & \vline\\
\hline
\vline& Equations of motion  & \vline& $\dot{M}=ad^*_{\Omega}M + mg\mathbf{\Gamma}\diamond\mathbf{\chi}$,  & \vline& $M_{k+1}=Ad^*_{\Omega_k^T}M_k +mg\mathbf{\Gamma}_k\diamond\mathbf{\chi}$, & \vline\\
\vline& & \vline& $\dot{\mathbf{\Gamma}}=-\Omega\mathbf{\Gamma}$  & \vline& $\mathbf{\Gamma}_{k+1}=\Omega_{k+1}\mathbf{\Gamma}_k$ & \vline\\
\hline
\vline&Right momentum map  & \vline& $J^R=P\diamond\Lambda +\mathbf{\Gamma}\diamond\mathbf{J}^{\Gamma}$
 & \vline& $J^R_{k+1}=P_k\diamond\Lambda_k + \mathbf{\Gamma}_{k}\diamond\tilde{\mathbf{J}}_{k}$ & \vline\\
\hline\\
\end{tabular}
\caption{Summary of the terms required to describe the \emph{body} representation of the \textbf{heavy top} in continuous and discrete time.}
\end{table}

\begin{table}[!ht]
\begin{tabular}{lllllll}
\hline
\vline & Property & \vline & Continuous &\vline & Discrete &\vline\\
\hline
\vline &Body attitude &\vline& $\Lambda(t)\in SO(N)$&\vline& $\Lambda_k\in SO(N)$ & \vline\\
\vline & Angular velocity &\vline&  $\omega=\dot{\Lambda}\Lambda^T=-\omega^T$ &\vline& $\omega_{k+1}=\Lambda_{k+1}\Lambda_{k}^T$ & \vline\\
\vline & Inertia Matrix &\vline& $I=\Lambda I_0 \Lambda^T$ &\vline& $I_k=\Lambda_k I_0 \Lambda_k^T$ & \vline\\
\vline & Angular mom. &\vline& $m=I\omega- \omega^T I$&\vline& $m_k=I_{k}\omega_k - \omega_k^T I_k$ & \vline\\
\vline & Position of c.o.m. &\vline& $\mathbf{\chi}=\Lambda\mathbf{\chi}_0$ &\vline& $\mathbf{\chi}_k=\Lambda_k\mathbf{\chi}_0$ & \vline\\ 
\hline
\vline & Eqns of motion &\vline& $\dot{m}=ad^*_{\omega}m - \nabla_I L \diamond I + mg\mathbf{\chi}\diamond \hat{\mathbf{z}}$, &\vline& $m_{k+1}=Ad^*_{\omega_{k}^T}m_k - \nabla_{I_k}L\diamond I_k
+mg\mathbf{\chi}_k\diamond \hat{\mathbf{z}}$, & \vline\\
\vline & &\vline& $\dot{I}=[\omega,I]$, &\vline& $I_{k+1}=\omega_{k+1}I_k\omega_{k+1}^T$, & \vline\\
\vline& &\vline& $\dot{\mathbf{\chi}}=\omega \mathbf{\chi}$ &\vline& $\mathbf{\chi}_{k+1}=\omega_{k+1}\mathbf{\chi}_k$
 & \vline\\
 \hline
\vline & Left mom. map &\vline& $J^L=\Lambda \diamond P -J\diamond I - \mathbf{J}^{\chi}\diamond\mathbf{\chi}$
&\vline& $J^L_{k+1}=\Lambda_k\diamond P_k+
G_k\diamond I_k + \tilde{\mathbf{J}}^{\chi}_{k}\diamond\mathbf{\chi}_{k}$ & \vline\\
\hline\\
\end{tabular}
\caption{Summary of the terms required to describe the \emph{spatial} representation of the \textbf{heavy top} in continuous and discrete time.}
\end{table}

\newpage

\begin{table}[!ht]
\begin{tabular}{lllllll}
\hline
\vline&Property  & \vline& Continuous  & \vline& Discrete & \vline\\
\hline
\vline&Body i attitude  & \vline& $\Lambda_i(t)\in SO(N)$ & \vline& $\Lambda_i^k\in SO(N)$  & \vline\\
\vline& Body i angular velocity  & \vline&  $\Omega=\Lambda_i^T\dot{\Lambda}_i=-\Omega_i^T$  & \vline& $\Omega^{k+1}_i=\Lambda^{k^T}_i\Lambda^{k+1}_i$ & \vline\\
 \vline & Position of c.o.m. of body i&\vline& $d_i$ &\vline& & \vline\\ 
 \vline &Orientat$^n$ matrix (rel. to b. 1) &\vline& $\Lambda=\Lambda_1^T\Lambda_2$ &\vline&  & \vline\\
\vline& Mod. inertia matrix of body i  & \vline& $\hat{I}_i=I_i -\frac{m_i^2}{m}D_{ii}$  & \vline& & \vline\\
\vline&Body 1 angular momentum  & \vline& $M_1=\hat{I}_1\Omega_1  - \Omega_1^T\hat{I}_1
-\epsilon\text{skew}(\Lambda \Omega_2 D_{12})$  & \vline& $M^k_1=\hat{I}_{1}\Omega^k_1 - \Omega^{k^T}_1 \hat{I}_1$  & \vline\\
 \vline&  & \vline& $$  & \vline& $+2\epsilon\text{skew}(D_{12}
 + \Omega_2^{k+1^T}D_{12}^T)$ & \vline\\
\vline&Body 2 angular momentum  & \vline& $M_2=\hat{I}_2\Omega_2  - \Omega_2^T\hat{I}_2
+\epsilon\text{skew}(D_{12}\Omega_1^T\Lambda)$  & \vline& $M^k_2=\hat{I}_{2}\Omega^k_2 - \Omega^{k^T}_2 \hat{I}_2$ & \vline\\
 
\vline&& \vline& $$  & \vline& $-2\epsilon\text{skew}(D_{12}
 - \Omega_1^{k+1^T}D_{12})$ & \vline\\
\hline 
\vline& Equations of motion  & \vline& $\dot{M}_i=ad^*_{\Omega_i}M_i$,  & \vline& $M_i^{k+1}=Ad^*_{\Omega_i^{k^T}} M_i^k$, & \vline\\
\vline& & \vline& $\dot{\Lambda}=\Lambda\Omega_2 -\Omega_1\Lambda$  & \vline& $\Lambda^{k+1}=\Omega_1^{{k+1}^T}\Lambda^k\Omega_2^{k+1} $ & \vline\\
\hline
\vline&Right momentum map  & \vline& $J^i_R=P_i\diamond\Lambda_i
+ J\diamond\Lambda$
 & \vline& $J_R^{i^k}=P^k_i\diamond \Lambda^k_i+ J^k\diamond\Lambda^k$ & \vline\\
\hline\\
\end{tabular}
\caption{Summary of the terms required to describe the \emph{body} representation of the \textbf{coupled rigid body} in continuous and discrete time, where $i\in\{1,2\}$.}
\end{table}

\begin{table}[!ht]
\begin{tabular}{lllllll}
\hline
\vline & Property & \vline & Continuous &\vline & Discrete &\vline\\
\hline
\vline &Body attitude &\vline& $\Lambda(t)\in SO(N)$&\vline& $\Lambda_k\in SO(N)$ & \vline\\
\vline & Angular velocity &\vline&  $\omega=\dot{\Lambda}\Lambda^T=-\omega^T$ &\vline& $\omega_{k+1}=\Lambda_{k+1}\Lambda_{k}^T$ & \vline\\
\vline & Inertia matrix &\vline& $\hat{I}_i=\Lambda_i\hat{I}_i^0\Lambda_i^T$ &\vline& $\hat{I}^k_i=\Lambda_i^k\hat{I}_i^0\Lambda_i^{k^T}$ & \vline\\
\vline & C.o.m. matrix &\vline& $D_{12}=\Lambda_1D_{12}^0\Lambda_2^T$ &\vline&$D^k_{12}=\Lambda_1^kD_{12}^0\Lambda_2^{k^T}$& \vline\\
\vline &  Sp.a.mom.(B1)&\vline& $m_1=\omega_1 \hat{I}_1 - \hat{I}_1\omega_1^T
$&\vline& $m^k_1=2skew(\hat{I}_1^k\omega_1^{k+1} -\epsilon \overbrace{\omega_1^{k+1^T}(\omega_2^{k+1}-I_d)D_{12}^{k^T}}^{\Gamma_1^{k+1}})$ & \vline\\
\vline &  &\vline& $\qquad +\epsilon\text{skew}(D_{12}\omega_2^T)$&\vline&& \vline\\
\vline &  Sp.a.mom.(B2)&\vline& $m_2=\omega_2 \hat{I}_2 - \hat{I}_2\omega_2^T
$&\vline& $m^k_2=2skew(\hat{I}_2^k\omega_2^{k+1} -\epsilon\overbrace{\omega_2^{k+1^T}(\omega_1^{k+1}-I_d)D_{12}^{k}}^{\Gamma_2^{k+1}})$ & \vline\\
\vline &  &\vline& $\qquad-\epsilon\text{skew}(\omega_1 D_{12})$&\vline&& \vline\\
\hline
\vline & Eqns of motion &\vline& $\dot{m}=\sum_{i} ad^*_{\omega_i}m_i + \nabla_{\hat{I}_i} L\diamond\hat{I}_i,$ &\vline& $m^{k+1}=\sum_{i} Ad^*_{\omega^{k^T}_i}m^{k+1}_i + \nabla_{\hat{I}^k_i}L\diamond\hat{I}^k_i,$ & \vline\\

\vline & &\vline& $\qquad +\nabla_{D_{12}} L\diamond D_{12}^T$, &\vline&  & \vline\\

\vline & &\vline& $\dot{\hat{I}}_i=[\omega_i,\hat{I}_i]$, &\vline& $\hat{I}_i^{k+1}=\omega_i^{k+1}\hat{I}^k_i\omega_i^{k+1^T}$, & \vline\\
\vline& &\vline& $\dot{D}_{12} =\omega_1D_{12} -D_{12}\omega_2
$ &\vline& $D_{12}^{k+1}=\omega_1^{k+1}D_{12}^k\omega_2^{k+1^T}$
 & \vline\\
 \hline
\vline & Left mom. map &\vline& $J^L=\sum_{i}P_i\diamond \Lambda_i  +J_i\diamond\hat{I}_i$
&\vline& $J^L=\sum_{i} P^k_i\diamond \Lambda^k_i + G^k_i\diamond \hat{I}^k_i
+ J^k\diamond D_{21}^k$ & \vline\\ 
 \vline &  &\vline& $\qquad + J \diamond D_{21}$&\vline&& \vline\\
\hline\\
\end{tabular}
\caption{Summary of the terms required to describe the \emph{spatial} representation of the \textbf{coupled rigid body} in continuous and discrete time, where $i\in\{1,2\}$.}
\end{table}

\end{small}
\newpage
\section{The Spatial DMV Algorithm for the Rigid Body}

For completeness, we include the specification of the (explicit) DMV algorithm for the spatial representation of the rigid body. Following the
approach of \cite{MCLACHLAN05}, the algorithm
uses the Schur decomposition of the Hamiltonian for the algebraic Ricatti
equation to construct a symmetric matrix $S_k$. Numerical experiments in
section \ref{sect:num_exp} show that there is little difference between the
conservative properties of the spatial and body versions of this algorithm.
We observe that the numerical round-off error differs between the two different
versions of the DMV algorithm depending upon the principal
moments of inertia. Numerical experiments, not presented in this
paper,  find negligible difference in the stability and computational performance between the two versions. 

\begin{enumerate}
\item For~~ $k=1:\rightarrow NT$\\
\item $m_{k}=\omega_{k-1} m_{k-1} \omega_{k-1}^T +[I_{k-1},\omega_{k-1}+\omega_{k-1}^T]$\\

\item $\mathcal{H}_k=\left( \begin{array}{cc}
\frac{m_k}{2}, &I_d\\ 
(\frac{m_k}{2})^2, & I_{k-1}^2 -\frac{m_k^T}{2}\\
\end{array}
\right)$\\

\item $[R_k,U_k]=\text{Schur}(\mathcal{H}_k)$\\
\item $S_k=(R_k)_{21}(R_k)_{11}^{-1}$\\
\item $\omega_k=(S_k+ \frac{m_k}{2})I_k^{-1}$\\
\item $I_{k}=\omega_k I_{k-1} \omega_k^T$\\
\item $k=k+1$\\

\end{enumerate}

\bibliographystyle{amsplain}

\end{document}